# Using $^{81}$Kr and Noble Gases to Characterize and Date Groundwater and Brines in the Baltic Artesian Basin on the One-Million-Year Timescale

Christoph Gerber[1]*; Rein Vaikmäe[2]; Werner Aeschbach[3]; Alise Babre[4]; Wei Jiang[5 *]; Markus Leuenberger[1]; Zheng-Tian Lu[5,6 *]; Robert Mokrik[7]; Peter Müller[5]; Valle Raidla[2]; Tomas Saks[4]; H. Niklaus Waber[8]; Therese Weissbach[3]; Jake C. Zappala[5,6]; Roland Purtschert[1]

[1]*Climate and Environmental Physics, Physics Institute and Oeschger Centre for Climate Change Research, University of Bern, 3012 Bern, Switzerland*

[2]*Tallinn University of Technology, Institute of Geology, 19086 Tallinn, Estonia*

[3]*Institute of Environmental Physics, University of Heidelberg, 69120 Heidelberg, Germany*

[4]*Department of Geography and Earth Sciences, University of Latvia, Riga, Latvia*

[5]*Physics Division, Argonne National Laboratory, Argonne, Illinois 60439, USA*

[6]*Department of Physics and Enrico Fermi Institute, University of Chicago, Chicago, Illinois 60637, USA*

[7]*Department of Hydrogeology and Engineering Geology of Vilnius University, LT-2009 Vilnius, Lithuania*

[8]*Institute of Geological Sciences, University of Bern, 3012 Bern, Switzerland*

**Current address: The University of Science and Technology of China, Hefei, Anhui 230026 China*

*Corresponding author:

cgerber@climate.unibe.ch

Climate and Environmental Physics, Physics Institute, University of Bern, Sidlerstrasse 5, 3012 Bern, Switzerland

# Abstract

Analyses for $^{81}Kr$ and noble gases on groundwater from the deepest aquifer system of the Baltic Artesian Basin (BAB) were performed to determine groundwater ages and uncover the flow dynamics of the system on a timescale of several hundred thousand years. We find that the system is controlled by mixing of three distinct water masses: Interglacial or recent meteoric water ($\delta^{18}O \approx -10.4‰$) with a poorly evolved chemical and noble gas signature, glacial meltwater ($\delta^{18}O \leq -18$ ‰) with elevated noble gas concentrations, and an old, high-salinity brine component ($\delta^{18}O \geq -4.5‰$, $\geq 90$ g $Cl^-$/L) with strongly depleted atmospheric noble gas concentrations. The $^{81}Kr$ measurements are interpreted within this mixing framework to estimate the age of the end-members. Deconvoluted $^{81}Kr$ ages range from 300 ka to 1.3 Ma for interglacial or recent meteoric water and glacial meltwater. For the brine component, ages exceed the dating range of the ATTA-3 instrument of 1.3 Ma. The radiogenic noble gas components $^{4}He^*$ and $^{40}Ar^*$ are less conclusive but also support an age of > 1 Ma for the brine. Based on the chemical and noble gas concentrations and the dating results, we conclude that the brine originates from evaporated seawater that has been modified by later water-rock interaction. As the obtained tracer ages cover several glacial cycles, we discuss the impact of the glacial cycles on flow patterns in the studied aquifer system.

# 1. Introduction

The evolution of brines and their mixing with – and replacement by – fresh water following changes in hydrogeological conditions have been intensely studied. For many sedimentary basins at high latitudes in North America and the crystalline Canadian Shield, the buildup and demise of Pleistocene ice sheets resulted in a considerable reorganization of groundwater flow systems (e.g. Person et al., 2007; Stotler et al., 2012). Shallow aquifers were flushed completely with glacial meltwater (Clark et al., 2000; McIntosh et al., 2012) and, in many places, relict meltwater is still found today (e.g. Person et al., 2007; McIntosh et al., 2012). In Europe, several hydraulic modeling and hydrogeochemical studies examined the effect of the Scandinavian ice sheet on groundwater dynamics (Voss and Andersson, 1993; Smellie et al., 1995; Boulton et al., 1995; 1996; Piotrowski, 1997; Laaksoharju et al., 1999; 2008; Zuzevicius, 2010; Saks et al., 2012). In the sedimentary Baltic Artesian Basin (BAB), located south of the Fennoscandian shield, glacial meltwater was identified (e.g. Vaikmäe et al., 2001; Raidla et al., 2009) and glacial flushing was studied by Zuzevicius et al. (2010) and Saks et al. (2012). Knowing how the Pleistocene glaciations influenced the flow dynamics in the BAB has important implications for sustainable management of the groundwater as a source of drinking water (Vaikmäe et al., 2001) and for potential long-term storage of radioactive waste or $CO_2$ (Shogenova et al., 2011).

In the deeper parts of the BAB, a high-salinity brine is present (Mokrik, 1997), the evolution of which has never been studied in detail. Many such types of brines originated from seawater, which was later modified by water-rock interaction. Proposed processes for the mineralization of such brines include evaporation (evaporative brines) and freezing (cryogenic brines) of seawater (e.g. Herut et al., 1990; Bottomley et al., 1999; Starinsky and Katz, 2003), besides the dissolution of evaporite rocks, or a combination of these (e.g. Knauth, 1988; Fontes and Matray, 1993; Fritz, 1997; Vengosh et al., 2000; Pinti et al., 2011; Millot et al., 2011; Bagheri et al., 2014). These processes are often distinguished based on the chemical composition of the brine, which is complicated by the dissolution and precipitation of rock forming minerals that further modify the chemical composition of the brine on different time scales (e.g. relatively fast reacting carbonate and sulphate mineral phases versus slowly reacting Al-silicate mineral phases). Recently, Greene et al. (2008) proposed using heavy noble gas concentrations as an alternative tool to differentiate between evaporation and freezing. The two processes result in very distinct changes in noble gas concentrations as salinity increases during brine formation. Far fewer processes can potentially change the noble gas concentrations after the brine formation compared to the range of processes that might change the chemical composition of the brine.

In this study, we focus on the Cambrian Aquifer System (CAS) of the BAB. This aquifer system is found across the entire basin and has been studied extensively in Northern Estonia, where it occurs close to the surface (Vaikmäe et al., 2001; Raidla et al., 2009; 2012; 2014). Raidla et al. (2009) proposed that groundwater in the northeastern part of the CAS represents a mixture of three distinct water types: a) recent meteoric fresh water, b) fresh glacial meltwater with an isotopic composition depleted in $^{18}O$ and $^{2}H$ ($\delta^{18}O$ as low as −23‰), and c) a brine component enriched in $^{18}O$ and $^{2}H$. Raidla et al. (2012) used $^{14}C$ to date the meltwater intrusion in Northern Estonia to 14–27 ka before present (BP), which is also the time of the advance and maximum extent of the Weichselian glaciation in the area (Kalm, 2012). Similarly, water in the overlying Middle-Upper Devonian aquifer system in Lithuania was also $^{14}C$-dated to a few thousand years to 27 ka ago (Mokrik et al., 2009). So far, the evolution and residence time of the deepest formation waters in the BAB with high brine fractions have only been qualitatively assessed by Mokrik (1997) based on the paleohydrogeological evolution of the area, the degree of the mineralization and relative ratios of major ions of the different groundwaters (e.g. $(Na^{+}+Cl^{-})/HCO_3^{-}$). The chemical composition of the deep formation waters (Mokrik, 1997) is considerably more evolved compared to that of groundwater from the Middle-Upper Devonian aquifer system (Mokrik et al., 2009), indicating residence times on the order of hundreds of thousands to millions of years.

Radioisotope dating of the groundwater in deeper (older) parts of the basin could further constrain the evolution of the brine and improve the understanding of present and past flow dynamics, but expected residence times are outside the dating range of $^{14}C$. Although $^{36}Cl$ would be applicable to the expected range of residence times, it would be difficult to interpret due to the unknown origin of the high salinity in the deeper part of the basin (Phillips, 2000). Similarly, $^{4}He$, $^{40}Ar$, and other radiogenic noble gas isotopes can be used for absolute dating only after calibration of their accumulation rates by independent age information (Torgersen and Stute, 2013). Under these conditions, $^{81}Kr$ is a near ideal tracer thanks to its chemical inertness, constant atmospheric concentration, and the likely absence of any significant subsurface sources or sinks other than radioactive decay (Collon et al., 2004, Jiang et al., 2012). Recent advances in the measurement technique resulting in higher counting efficiency and thus smaller sample size and shorter measurement time (Jiang et al., 2012) have enabled a more widespread application of this tracer in dating old groundwater (e.g. Sturchio et al., 2014; Aggarwal et al., 2015).

Here, we present chemical, isotope, and noble gas data of groundwater collected from the Cambrian aquifer system (CAS) of the BAB, with a special emphasis on $^{81}Kr$. Because ice sheets reorganized groundwater flow on a regional scale, the CAS should also be studied at a regional scale. Most of the past studies were conducted on more local scales, except for two modeling studies (Saks et al., 2012;

Virbulis et al., 2013), the latter of which estimated the hydraulic age of groundwater in the CAS to be on the order of several hundreds of ka to 1 Ma. In the light of such long residence times, it is crucial to consider the effect of repeated glacial cycles on the long-term evolution of groundwater composition and flow. Sampling the deeper parts of the CAS on a regional scale for chemistry, noble gases, and multiple dating tracers ($^{81}$Kr, $^{85}$Kr, $^{39}$Ar, $^{14}$C, $^{4}$He, $^{40}$Ar) allows us to elucidate the evolution of the brine, mixing proportions of the different groundwater components, and the flow dynamics over the last 1 Ma.

## 2. Study Area

The Baltic Artesian Basin (BAB), one of the largest artesian basins in Europe, covers the territory of Latvia, Lithuania and Estonia, parts of Poland, Russia, and Belarus as well as a large area of the Baltic Sea (Fig. 1a). Geologically, the BAB is situated in the northwestern part of the East European platform on the southern slope of the Fennoscandian Shield. The Proterozoic crystalline basement is covered by some Proterozoic (Ediacaran) and mainly Paleozoic (Cambrian to Cretaceous) sedimentary rocks. To the north, northwest, and southeast, the basin is delimited by the outcropping crystalline basement (Mokrik, 1997; Raukas and Teedumäe, 1997). From there, the basement and the overlying sediments gradually dip towards the central part of the BAB by 2–4 m/km (Fig. 1c). In the east and the southwest, the BAB borders the Moscow basin and the Danish-Polish basin, respectively. The southwest dipping sediments reach a maximum thickness of ~600 m in the east and of ~5000 m at the border to the Danish-Polish Basin (Fig. 1b; Mokrik, 1997).

Hydrogeologically, the BAB is a complex multi-layered system of aquifers and aquitards (Mokrik, 1997). The Cambrian aquifer system (CAS), which is the focus of this study, directly overlies the crystalline basement in most of the BAB (Fig. 1) and consists mainly of dolomitic sandstones and siltstones (Raidla et al., 2006). In the northeastern part, the system is also known as Cambrian-Vendian (Ediacaran) aquifer system (Cm–V) due to the occurrence of Ediacaran sandstones and clays. There, the CAS constitutes an important drinking water resource. The thickness of the CAS averages at about 80–100 m, but varies from about 50 m in the central part to 200 m at the northern and southern fringes to several 100 m near the Danish-Polish Basin (Poprawa et al., 1999; Luksevics et al., 2012). The CAS is covered by less permeable dolomitic limestones and claystones of the Ordovician-Silurian (O–S) formation (Fig. 1b, c) that reach a thickness of more than 1 km in the deeper parts of the BAB and pinch out in the southeastern and northern parts (Poprawa et al., 1999; Luksevics et al., 2012).

Both formations, the CAS and the O–S, have been faulted considerably during the Late Silurian and Early Devonian, especially along the Liepāja-Pskov fault zone in the central part of the BAB (cf. Fig. 1;

Raukas and Teedumäe, 1997; Brangulis and Kanevs, 2002). Later major tectonic activity was mostly limited to the glacioisostatic downwarping and uplift caused by the late Cenozoic glaciations (Mokrik, 1997; Raukas and Teedumäe, 1997). Faults are an important factor for aquifer connectivity, as they might either limit the groundwater flow or act as vertical conduits and connect otherwise disconnected aquifers (Virbulis et al., 2013). In addition, they may also have acted as potential pathway for the infiltration of the brine from the surface.

The present-day hydraulic heads indicate recharge of the CAS in the southeast along the Belarus-Masurian Anticline, a general northward flow direction, and discharge of groundwater to the Baltic Sea. Virbulis et al. (2013) estimated a hydraulic conductivity $K$ of 2.95 m/day for the CAS. In contrast, for the thick overlying O–S rock formation, they obtained a hydraulic conductivity on the order of $10^{-9}$ m/day, making it an effective aquitard. This situation results in a predominately horizontal groundwater flow in the CAS (Virbulis et al., 2013). Several authors have suggested that regional groundwater flow in the BAB, including the CAS, reversed during the various glacial cycles of the Quaternary (Jõeleht, 1998; Vaikmäe et al., 2001; Vaikmäe et al., 2008; Zuzevicius, 2010; Saks et al., 2012), when the Scandinavian ice sheet covered much of the BAB (cf. Fig. 1a; Guobyte and Satkunas, 2011; Kalm, 2012).

## 3. Methods

### 3.1. Sampling and Analytical Methods

Groundwater samples were collected from seven deep wells (cf. Table 1, Fig. 1) during three field campaigns: (I) from May to September 2012 for $^{14}$C analyses, (II) in October 2012 for $^{81}$Kr, $^{39}$Ar, $\delta^{18}$O, $\delta^{2}$H, and chemical analyses, and (III) from June to October 2013 for noble gas analyses. To monitor air contamination during gas extraction for $^{81}$Kr and $^{39}$Ar dating and/or inter-aquifer leakage in the well, $^{85}$Kr analyses were conducted. More details on the fieldwork and the correction for (air) contamination are given in the Electronic Appendix A.

Radiocarbon samples were extracted by direct precipitation of dissolved inorganic carbon (DIC) from 200–300 L of water (Aggarwal et al., 2014). Radiocarbon activities in groundwater were measured by ultra-low-level liquid scintillation (Quantulus 1220) at Tallinn University of Technology. The $^{14}$C results are reported as percentage of modern carbon (pmC) with a measurement error of the radiocarbon activity of ±0.5 pmC.

For $^{81}$Kr and $^{39}$Ar dating, a sample size of 2000–3000 L of water is required. These large-volume samples were pumped through a vacuum extraction chamber and the extracted gas transferred to a steel container (Purtschert et al., 2013; Lu et al., 2014). Extraction of Kr and Ar from the bulk gas was performed at the Physics Institute, University of Bern, by preparative gas chromatography (Purtschert et al., 2013; Lu et al., 2014).

The isotope ratios $^{81}$Kr/Kr and $^{85}$Kr/Kr were determined using the ATTA-3 instrument in the Laboratory for Radiokrypton Dating, Argonne National Laboratory (Jiang et al., 2012). The Atom Trap Trace Analysis method (ATTA) is a selective and efficient atom counter capable of measuring both $^{81}$Kr/Kr and $^{85}$Kr/Kr ratios of environmental samples in the range of $10^{-14}$–$10^{-10}$ (Chen et al., 1999). For $^{81}$Kr dating in the age range of 50 ka–1.3 Ma, the required sample size is 5–10 µL (STP) of krypton gas, which can be extracted from approximately 100–200 L of water. Abundances of $^{81}$Kr are reported relative to modern atmospheric air as $R_{sample}$ = $(^{81}Kr/Kr)_{sample}/(^{81}Kr/Kr)_{air}$. For $^{85}$Kr, the results are reported in the conventional units of dpm/$cc_{Kr}$, which stands for the number of $^{85}$Kr decays per minute per mL (STP) of Kr gas (for conversion, 100 dpm/$cc_{Kr}$ corresponds to a $^{85}$Kr/Kr ratio of $3.03 \times 10^{-11}$).

Activities of $^{39}$Ar were measured by low-level gas proportional counting at the Physics Institute, University of Bern (Loosli, 1983; Loosli et al., 1986; Forster et al., 1992). Results are expressed in %modern Ar, relative to the atmospheric activity concentration of 1.78 mBq/$L_{Ar}$ (Loosli et al., 2000).

Stable isotope ratios, expressed in standard delta notation ($\delta^{18}O$ and $\delta^{2}H$), were measured at the Physics Institute, University of Bern, with a Picarro cavity ring-down laser analyzer. Duplicate measurements are generally in good agreement (±0.5‰ for $\delta^{18}O$ and ±1‰ for $\delta^{2}H$), close to the analytical uncertainty of 0.1‰ and <1‰ for $\delta^{18}O$ and $\delta^{2}H$, respectively.

The samples for chemical analysis were stored at room temperature (21°C) in brown 200 mL glass flasks with a rubber sealing until analysis at the Institute of Geological Sciences, University of Bern, in January 2015. The samples were filtered with 0.45 µm PES-filters. For the various analyses, the groundwater samples were gravimetrically diluted in up to four different dilutions (1:100 to 1:2000) due to the elevated salinity. Total inorganic carbon (TIC) and total organic carbon (TOC) were analyzed using a TIC/TOC Analyzer (Analytic Jena multi N/C 2100S). Major anions and cations were analyzed by ion-chromatography (Metrohm ProfIC AnCat MCS) and trace elements by ICP-OES (Varian 710ES). Chemical data from Northern Estonia were measured by ion-chromatography at the University of Vilnius. The analytical error of all the chemical analyses is ±5% with the total uncertainty being better than ±10%. Mineral saturation states were modeled with the geochemical code PHREEQC (Parkhurst and Appelo, 1999; v. 2.18) using the Wateq4f and Pitzer activity formulations and databases for samples with salinities below and above that of seawater, respectively.

For noble gas analysis, water samples were transferred to copper tubes in-line with the well, and subsequently the tubes were immediately pinched off (Beyerle et al., 2000; Aeschbach-Hertig and Solomon, 2012). The samples were analyzed with a GV Instruments MM5400 noble gas mass spectrometer at the Institute of Environmental Physics, Heidelberg (Freundt et al., 2013). Measurement precision is better than 1% for light noble gas concentrations, 1–2 % for the heavy noble gases, and better than 0.5 % for the Ne isotope ratio.

## 3.2. Evaluation of the Noble Gas Data

Often, groundwater is found to be oversaturated relative to air-equilibrated water (AEW) and the additional amount of dissolved gases is termed excess air (EA, Heaton and Vogel, 1981) and expressed as relative Ne excess ΔNe:

$$\Delta Ne = \left(\frac{C^{Ne}_{measured}}{C^{Ne}_{equi}(T,S)} - 1\right) \cdot 100\% \tag{1}$$

For a mixture of different water components, as in the BAB, two definitions of $C^{Ne}_{equi}(T,S)$ are possible: a) ${}_sC^{Ne}_{equi}$ – the concentration in air-equilibrated water at the salinity and temperature measured during sampling – which is meaningful for the assessment of the potential for degassing during sampling and b) the proportion-weighted air-equilibrated Ne concentrations of the – in this study three – mixing components:

$$_{mix}C^{Ne}_{equi} = \sum_{i=1}^{3} f_i \cdot \ {}^{i}c^{Ne}_{equi}\,(T_i, S_i) \tag{2}$$

where $f_i$ are the mixing proportions of the water components. This second definition, which requires some knowledge about the composition of the mixing components, reflects the amount of excess air added during infiltration.

Groundwaters oversaturated in gases at ambient conditions are prone to degassing, which may be a diffusion-controlled and/or solubility-controlled process, resulting in characteristic noble gas fractionation patterns (e.g. Aeschbach-Hertig et al., 2008; Aeschbach-Hertig and Solomon, 2012). While the groundwaters sampled in this study were all undersaturated in-situ, this may have changed into oversaturation during the ascent of the water in the wells due to the decreasing hydraulic pressure. The simplest form of fractionation is the Rayleigh fractionation model (Stute and Schlosser, 2000; Ballentine et al., 2002):

$$\frac{C_i}{C_i^0} = \left(\frac{C_{Ne}}{C_{Ne}^0}\right)^{\alpha_i(T)} = f^{\alpha_i(T)} \qquad \alpha_i^{sol}(T) := \frac{K_h^i(T)}{K_h^{Ne}(T)} \qquad \alpha_i^{diff}(T) := \left(\frac{D_i(T)}{D_{Ne}(T)}\right)^n \tag{3}$$

Where $C^0_{Ne}$ is the elemental concentration of Ne before degassing, $C_i^0$ the elemental concentration of another noble gas or of a minor Ne isotope before degassing, $C_i$ and $C_{Ne}$ the respective concentrations after degassing, $f$ is the fraction of remaining Ne, and $\alpha_i(T)$ the temperature dependent fractionation factor. For solubility-controlled degassing, the fractionation factor $\alpha_i^{sol}(T)$ is the ratio of the Henry constants $K_h$. This model is appropriate if initially noble gas-free gas bubbles are moving through a water column (Ballentine et al., 2002). For diffusion-controlled degassing, the fractionation factor $\alpha_i^{diff}(T)$ is the ratio of the diffusion coefficients *D* (as published by Jähne et al. (1987) and Bourg and Sposito (2008)) to the power of *n*. Models of gas transfer suggest 0.5, 2/3, and 1 as possible values for *n* with smaller numbers referring to increasingly turbulent systems (Holmen and Liss, 1984). Early excess-air degassing models used *n*=1 (e.g. Stute and Schlosser, 2000), however, recently it has been pointed out that n could be smaller (e.g. Aeschbach-Hertig et al., 2008; Aeschbach-Hertig and Solomon, 2012). For diffusion-controlled degassing, *f* can be calculated from the $^{20}$Ne/$^{22}$Ne isotope fractionation by rearranging Equation 3:

$$R_{Ne} = R_{Ne}^0 \cdot f^{\alpha-1} \qquad \alpha = \left(\frac{D_{20}}{D_{22}}\right)^{\frac{2}{3}} = 1.0141^{\frac{2}{3}} = 1.0094 \tag{4}$$

with $R_{Ne} \coloneqq C_{20Ne}/C_{22Ne}$ the measured isotopic ratio and $R_{Ne}^0 \coloneqq C_{20Ne}^0/C_{22Ne}^0$ the atmospheric ratio and the ratio of isotopic diffusion coefficients taken from Bourg and Sposito (2008).

## 3.3. Deriving Groundwater Ages

For easier reading, we employ the following definitions of $^{81}$Kr ages:

$^{81}$Kr age ($^{81}t$): Apparent groundwater residence time of a water sample that neglects dispersive mixing (also known as piston-flow age) and that is simply calculated by (Sturchio et al., 2004; Purtschert et al., 2013):

$$^{81}t = -1/\lambda_{81} \ln(R/R_{air}) \tag{5}$$

Where $\lambda_{81}$ = 3.03(±0.14) x $10^{-6}$ $yr^{-1}$ is the decay constant of $^{81}$Kr.

$^{81}Kr^m$ age ($^{81}t^m$): Calculated (deconvoluted) $^{81}$Kr age of a water component contributing to a given mixed water sample. These ages depend on the specific mixing proportions of each sample and are therefore marked with the superscript $^m$.

The individual end members – in this study interglacial or recent meteoric water (henceforth called meteoric water), glacial meltwater, and brine – may have had distinct noble gas signatures before mixing and likely had different $^{81}Kr^m$ ages. For a detailed interpretation of the measured $^{81}$Kr activities, it is therefore necessary to separate the measured noble gas concentrations to obtain the contribution from each end member. Thereby, two effects have to be considered: a) the nonlinearity of decay ages and b) the different Kr concentrations of the mixing endmembers.

a) The (apparent) $^{81}$Kr age underestimates the end-member-proportion-weighted age of a mixture $t_{mix}$ due to the nonlinear variation of $^{81}$Kr with age (Suckow et al., 2013):

$$t_{mix} = \sum_{i=1}^{3} {}^{81}t_i^m \cdot f_i^{H2O} = -\frac{1}{\lambda_{81}} \sum_{i=1}^{3} \ln R_i \cdot f_i^{H2O} \geq -\frac{1}{\lambda_{81}} \ln R_{mix} = {}^{81}t \qquad (6)$$

where $f_i^{H2O}$ is the proportion of water from end member *i*.

b) The $^{81}$Kr content of the mixture ($R_{mix}$) is the sum of the Kr concentration-weighted $^{81}$Kr content of each end member ($R_i$):

$$R_{mix} = \sum_{i=1}^{3} f_i^{Kr} \cdot R_i \qquad \text{with } f_i^{Kr} = C_i^{Kr} \cdot f_i^{H2O} \cdot n \qquad (7)$$

where $f_i^{H2O}$ and $f_i^{Kr}$ are the water and Kr proportions of each end member of sample *i* and *n* is a normalization constant. This distinction between $f_i^{H2O}$ and $f_i^{Kr}$ is often neglected because it is assumed that all $C_i^{Kr}$ are similar.

For accumulating gases, such as $^{4}$He and $^{40}$Ar, the concentration is a linear function of age and therefore the measured (mean) concentration $\bar{c}$ corresponds to the mean age of the mixed sample $\bar{t}$:

$$\bar{t} = \sum_{i=1}^{3} t_i \cdot f_i^{H2O} = \sum_{i=1}^{3} C_i / r \cdot f_i^{H2O} = \frac{1}{r} \sum_{i=1}^{3} C_i \cdot f_i^{H2O} = \frac{1}{r} \bar{C} \qquad (8)$$

Furthermore, also correction b) does not apply to these tracers, as they are reported as absolute concentrations rather than isotopic ratios.

# 4. Results

## 4.1. Chemical Composition

Groundwaters show a considerable range in chemical composition (Table 2). Groundwater pH values measured in the field range between 7.0 and 8.1 (Table 2) and become 5.75 and 7.20 when adjusted for calcite saturation (see section 5.1.3). At the well head, mildly reducing redox potentials prevail ($Eh_{Ag/AgCl}$ –72 to –12 mV) in agreement with observed concentrations of $CH_4$ (Table 3) and the absence of measurable $NO_3^-$ (<1.6 mg/L; Table 2). Total dissolved solids (TDS) range from brackish water (5.1 g/kg) to brines that are enriched in salt by 3–5 times relative to seawater (up to 144 g/kg). Groundwaters are of the general Na-Cl type, with elevated amounts of $Ca^{2+}$ or $Mg^{2+}$ in the more saline groundwaters. Ionic concentrations display linear relationships between $Cl^-$ and $Na^+$, $K^+$, $Li^+$, $Br^-$

, and $SO_4^{-2}$ (except well 6) from the brackish to the saline groundwaters. Molar ratios of $Na^+/Cl^-$ and $Br^-/Cl^-$ are close to those of seawater. The absolute concentrations of $Cl^-$, $Na^+$, and $Br^-$ extend the range of previously collected groundwater from the CAS along the line defined by evaporation of seawater and dilution of seawater with pure water (Fig. 2).

## 4.2. Stable Isotopes of Water ($\delta^{18}O$ and $\delta^2H$)

Stable isotopes of the water molecule range from −13.6‰ to −4.4‰ for $\delta^{18}O$ and from −101‰ to −35‰ for $\delta^2H$ (Fig. 3, Table 2) and vary in their D-excess ($D=\delta^2H-8*\delta^{18}O$), indicating a varying origin of the dominant water component in the groundwater samples. Groundwater from wells 1 and 2 are isotopically depleted in $^{18}O$ and $^2H$ compared to the present-day annual-mean isotopic composition of −10.6 ‰ for $\delta^{18}O$ in Riga (IAEA/WMO, 2016). In contrast, the saline samples are enriched in $^{18}O$ and $^2H$ compared to the present-day annual mean. These shifts in isotopic composition are explained by mixing of meteoric fresh water with isotopically depleted (but meteoric) glacial meltwater and an isotopically enriched brine component, as proposed by Raidla et al. (2009). The shift in isotopic composition away from the MWL for the more saline samples suggests a non-meteoric source of the brine component or evaporation and/or high temperature water-rock interaction (Horita, 2005).

## 4.3. Dissolved Gases

Noble gas data from campaign III are shown in Table 3, including, for comparison, air-equilibrated concentrations for the possible water components. The atmospheric noble gas concentrations vary quite widely, ranging from significantly above to far below air-equilibrated fresh water with increasing salinity. Groundwater samples from wells 1, 2, 5, and 7* are considerably oversaturated in Ne with respect to both definitions of $C_{equi}^{Ne}$, especially for wells 1 and 2 with the highest glacial meltwater proportion (Table 3). Shallower and less mineralized groundwater samples from the CAS in Northern Estonia exhibit even higher amounts of excess air (Weißbach, 2014), which is attributed to the recharge of glacial meltwater either through basal melting (Vaikmäe et al., 2001) or infiltration of surface meltwaters (Grundl et al., 2013). In contrast, negative $\Delta Ne_s$ for wells 3 and 4 indicate degassing. A similar pattern is also observed for the heavier noble gases.

All samples contain several orders of magnitude more $^4He$ than does AEW (Table 3). $^3He/^4He$ ratios are depleted compared to the atmosphere ($^3He_{atm}/^4He_{atm} = 1.38\cdot10^{-6}$) and are typical for the crustal production ratio, which is on the order of $2\cdot10^{-8}$ (Mamyrin and Tolstikhin, 1984; Torgersen and Stute, 2013). The $^{40}Ar/^{36}Ar$ ratios (Table 3) are considerably higher (up to 419) than the atmospheric ratio of $R_{atm}^{Ar}$ = 298.6 (Lee et al., 2006). Both He and Ar thus indicate the presence of a non-atmospheric component.

The concentrations of major gases inferred from the gas extracted in the field for $^{81}Kr$ analysis (campaign II) complement the noble gas data from campaign III (Table 3). The Ar concentrations from the two sampling campaigns are generally consistent with each other for an extraction yield of 80–90% of the field degassing (see Fig. A2 in the Electronic Appendix C). Total dissolved gas contents of the groundwater range from 19–107 $cm^3$ (STP)/kg for an extraction yield of 85% (Table 3). The composition of the extracted gas is dominated by $N_2$ (>97%), with the exception of well 6, where $CH_4$ constitutes 60% of the gas. $N_2$ concentrations in the groundwater correspond to a dissolved gas pressure of about 2.5–11 bar in the samples, based on $N_2$ solubility extrapolated from seawater (Battino et al., 1984).

Oversaturation of $N_2$ relative to AEW is considerably higher than noble gas oversaturation, suggesting subsurface $N_2$ sources. Measured redox potentials lie within those required for denitrification and $N_2$ reduction, the latter in agreement with the absence of measurable $NH_4^+$. The similarly absent $NO_3^-$ suggests that denitrification might have produced some $N_2$. However, without any support from isotope and bacteriological data, it is not possible to judge to what degree denitrification might explain the excess in $N_2$. Similarly unknown is the original source of $CH_4$ except that the isotopic composition indicates a biogenic origin (see Raidla et al., 2012 for discussion). Based on measured redox potentials and isotope compositions of dissolved $CH_4$, TIC, $SO_4^{2-}$ (Raidla et al., 2012, 2014), it appears that, currently, $CH_4$ in the deep CAS groundwaters is being consumed by sulphate reduction rather than being produced by methanogenesis.

## 4.4. Radioisotopes

The results of $^{85}Kr$, $^{39}Ar$, $^{14}C$, and $^{81}Kr$ analyses are given in Table 4. The $^{85}Kr$ concentrations for samples 2, 4, and 6 are likely the result of contamination either during gas extraction or in the well through a leaky casing. The estimated contamination is 2–13% and $^{39}Ar$ and $^{81}Kr$ activities were corrected accordingly (Table 4, see Electronic Appendix A for details of the contamination correction). After this correction, only well 1 and, possibly, well 4 exhibit remaining $^{39}Ar$ activity. For $^{14}C$ of DIC, no correction was applied, because $^{14}C$ was sampled during a separate sampling campaign and because the degree of contamination could be different for gases and dissolved constituents.

Contamination-corrected isotopic abundances of $^{81}Kr$ relative to modern atmospheric air ($R/R_{air}$) range from 0.00±0.02 to 0.38±0.03. This constrains the upper level of subsurface-produced $^{81}Kr$ to 0.02 $R/R_{air}$ in the CAS, confirming a previous estimate by Lehmann et al. (1993) that underground production is negligible for $^{81}Kr$ in most settings. Consequentially, underground production is neglected in the $^{81}Kr$ age calculations in this article. Abundances of $^{85}Kr$, $^{39}Ar$, $^{14}C$, and $^{81}Kr$ below detection limits indicate the absence of water younger than 50 a, 1000 a, 35 ka, and 1.3 Ma, respectively. The presence of measurable $^{39}Ar$ (well 1 and 4) or $^{14}C$ (well 4) activities in significantly $^{81}Kr$ depleted waters is likely due to mixing of water components of different age. For example, Well 1 could be explained by a binary mixture of tracer-free very old water with a water component that is a few hundred years old ($^{85}Kr$- but not $^{39}Ar$-free). The low $^{14}C$ value in well 1 probably is the result of addition of $^{14}C$-dead carbon, likely from the carbonate cement present in the aquifer (Raidla et al., 2006; 2012). The reason for the relatively high $^{14}C$ activity of well 4, compared to the low $^{81}Kr$ value, is unclear, but could be related to contamination, since $^{14}C$ was not contamination corrected ($^{39}Ar$ is within 2 $\sigma$< DL). In the CAS, underground production of $^{39}Ar$ is negligible because $^{39}Ar$ activities are below the detection limit for several wells.

Based on the flow regime indicated by the present-day hydraulic head distribution (Recharge in the southeast, flow to the northwest and finally discharge in the north, Fig. 4, Virbulis et al., 2013), we expect relatively high $^{81}Kr$ abundances in the southeast and decreasing values towards the Baltic Sea. Fig. 4 shows that the well with the highest $^{81}Kr$ abundance is indeed located closest to today's recharge area in the South (sample 5). However, the $^{81}Kr$ abundances of the northernmost samples (1 and 2) are relatively high as well, in contradiction with the expected spatial pattern.

# 5. Discussion

## 5.1. Genesis of the Brine

### 5.1.1. Origin of the Brine

The origin, time of formation, and chemical evolution of brines in sedimentary basins and crystalline rock environments are still under debate (e.g. McNutt et al., 1987; Knauth, 1988; Herut et al., 1990; Frape et al., 2003; Starinsky and Katz, 2003; Leybourne and Goodfellow, 2007; Greene et al., 2008; Stotler et al., 2012; Bagheri et al., 2014). With respect to the origin and the time of formation of deep-seated brines, different scenarios have been invoked. These include dissolution of evaporite rocks, evaporation of seawater (with or without later modification by water-rock interaction), and enrichment of seawater by cryogenic or ion-filtration processes. All these processes carry certain qualitative information about the brine formation time and need to be consistent with the paleo-hydrogeological evolution of the basin. Evaporite dissolution may take place whenever groundwater is exposed to evaporitic sediments. Evaporative enrichment of seawater requires the presence of seawater and appropriate climatic and surface hydraulic conditions such that evaporation in the basin is larger than precipitation and other inflows to the basin. Such conditions are unlikely to have existed in the Baltic Basin during the Quaternary (e.g. Wohlfahrt et al., 2008; Kalm et al., 2011; Guobyte and Satkunas, 2011) and evaporative brine formation must therefore date back to Tertiary times or earlier (> 2.6 Ma ago) when more suitable conditions prevailed in the basin (Jansen et al., 2007). Ion-filtration processes are most efficient during burial of the sediment pile and are thus related to the diagenetic phase(s) of the sedimentary basin. Therefore, ion-filtration processes also commonly date back to pre-Quaternary times, but may have been re-induced during the Quaternary by the glacial overburden. Cryogenic enrichment of brackish water or seawater requires a cold climate and might therefore have occurred during Quaternary times.

The chemical composition of deep-seated groundwater in the CAS is most consistent with an evaporative origin of the brine. An origin of the brine purely from dissolution of halite-bearing evaporite rocks is incompatible with observed $Na^+/Cl^-$ and $Br^-/Cl^-$ ratios (cf. Fig. 2). Furthermore, evaporitic sequences are known only from the Permian sediments but have not been found in the CAS (Kalvāns, 2012). Some impacts of mineral dissolution/precipitation reactions including evaporite dissolution are, however, present, as indicated by the ratios of $K^+/Cl^-$, $Li^+/Cl^-$, which deviate in opposite directions from the seawater evaporation line (cf. Fig. 2) and the almost constant $(Ca^{2+}+Mg^{2+})/Cl^-$ ratio with a well-developed linear correlation between $Ca^{2+}$ and $Mg^{2+}$. Furthermore, the various mineral saturation states (Table 5) all indicate carbonate, sulphate, and (Al-)silicate mineral dissolution and precipitation in addition to evaporative enrichment of seawater. Cryogenic formation of saline groundwaters has been documented for crystalline rocks in the Baltic region (Bein and Arad, 1992). However, cryogenic enrichment is incompatible with the observed $\delta^{18}O$–$\delta^{2}H$ signature of the saline groundwaters (Fig. 3) as it would shift the isotope signature of the residual water to the left of the GMWL (e.g. Stotler et al., 2012). Evaporation shifts it to the right of the GMWL, as observed (Fig. 3). Furthermore, cryogenic enrichment of seawater and ion-filtration during diagenesis seem incompatible with the observed chemistry in the sedimentary CAS, as these would result in different fractionation patterns of dissolved components.

Different brine formation processes also lead to distinct functional dependencies between noble gas concentrations and salinity as shown in Fig. 5 (Greene et al., 2008). In the case of ion-filtration (and evaporite dissolution), the initial noble gas concentrations of the seawater would be mainly conserved, because the clay acts as a membrane that impedes the passage of ions, but may allow the neutral noble gases to pass through together with the water (Appelo and Postma, 1993). Evaporating brines remain in contact with the atmosphere during their genesis so that concentrations decrease in parallel to the decreasing solubility due to increasing salinity, resulting in low noble gas concentrations. For cryogenic brine formation on the other hand, the overlying ice impedes contact with the atmosphere. This isolation results in highly enriched heavy noble gas concentrations in the residual water, because they cannot be incorporated into the ice lattice, whereas the lighter isotopes (He, Ne) can be incorporated and become depleted in the residual water (Hood et al., 1998; Greene et al., 2008; Malone et al., 2010). Thus, the decreasing noble gas concentrations with increasing salinity clearly support an evaporative origin (Fig. 5). In conclusion, based on chemical, stable isotope, and noble gas signatures of the most saline groundwaters, the here preferred origin of the brine in the BAB is one of evaporated seawater that has later been modified by water-rock interaction and diluted by fresh water including glacial meltwater.

### 5.1.2. Chemical Modification of the Brine

Considering the aquifer mineralogy (Raidla et al., 2006) and the long timescales, in-situ, all groundwaters are in equilibrium with respect to calcite and dolomite. Modeling of the mineral saturation states shows that at calcite saturation, the brine-dominated groundwaters (wells 3 to 7*) are in equilibrium with gypsum and celestite, except for the most saline groundwater from well 6 (Genciai). The calculated equilibrium with respect to barite and quartz (Table 5) indicates an evolved geochemical evolution for the brackish groundwaters and, based on mineral dissolution kinetics, suggests a considerable, albeit significantly shorter, residence time of the groundwater mixture compared to the brine-dominated groundwater mixtures.

The modeled equilibrium state of gypsum and celestite (Table 5) for samples with a high brine proportion cannot be explained by dissolution of these minerals within the aquifer itself, as gypsum and celestite are not observed in the rocks of the CAS (Raidla et al., 2006). Alternative explanations include a brine component that had been equilibrated with gypsum and celestite prior to infiltration into the CAS and/or sulfate reduction in the groundwater within the aquifer. For the latter there is ample evidence from S- and C-isotope investigations on shallower, less mineralized groundwater from the CAS (Raidla et al., 2012, 2014) and – for the saline deep groundwaters – also from well 6, for which the low $SO_4^{-2}$ and high organic carbon (TOC) contents (Table 2) indicate sulfate reduction. Highly elevated $CH_4$ concentrations in well 6 indicate microbial methanogenesis in addition to sulfate reduction. Such substantial sulfate reduction results in strong undersaturation of this highest mineralized groundwater with respect to $SO_4$-bearing mineral phases (Table 5). This suggests that in the bulk of brine-dominated groundwaters, gypsum saturation was reached before infiltration of the brine component and later dilution with fresh water types was not efficient enough to deviate the groundwater composition from this equilibrium. In line with this is the modelled undersaturation with respect to gypsum and celestite in the brackish groundwaters (wells 1 and 2) with a) much higher proportions of freshwater components and b) potential ongoing sulfate reduction as observed in shallower groundwater of the CAS (Raidla et al., 2014).

A mixing of fresh water, glacial meltwater, and brine components has been observed in other sedimentary basins mainly in North America (e.g. Pearson et al., 1991; McIntosh and Walter, 2006; McIntosh et al., 2012), but also for the CAS in Northern Estonia (Raidla et al., 2009). The patterns of chemical composition, saturation states, and stable isotopes for samples from the deeper parts of the CAS all indicate a common evolution of the collected groundwaters by gradual dilution of a single brine end member by fresh water including glacial meltwater.

### 5.1.3. Degassing

The low noble gas concentrations of wells 3 and 4 indicate the occurrence of degassing (Fig. 5). Depleted $^{20}Ne/^{22}Ne$ ratios compared to the atmospheric value ($R_{atm}^{Ne} = 9.80$, Eberhardt et al., 1965) for these wells point to diffusion-controlled degassing (Peeters et al., 2002). Diffusive degassing is likely driven by gas stripping into a gas phase depleted in noble gases (e.g. $N_2$ or $CH_4$, similar to Sültenfuss et al., 2011) in accordance with the negative $\Delta Ne_s$ values of wells 3 and 4 (Table 3). An additional complication in our case is the presence of mixing of very distinct water components that may have individual degassing histories. Concerning the timing of the degassing, three scenarios are possible: a) gas loss during transfer to or storage in the Cu tubes; b) gas stripping during the rise of the groundwater in the borehole due to decompression (Late degassing); and c) gas stripping during infiltration of the brine (Early degassing, before mixing) caused for example by gases released by tectonic activities or geochemical reactions such as denitrification and methanogenesis in the recharge zone at that time. Option a) is unlikely because Ar concentrations determined from sampling campaigns II and III generally show consistent results (see Fig. A2 in the Electronic Appendix C).

Gas stripping in the groundwater body due to decompression and ebullition of oversaturated $N_2$ is a well-known mechanism and cannot be excluded. Such a late degassing process would also cause a loss of $CO_2$, explaining the apparent oversaturation of all sampled groundwaters with respect to calcite when using the pH value measured in the field. However, the fact that the partial pressure of $CO_2$ in the samples calculated with the field pH value is more than one order of magnitude higher than that of the atmosphere (Table 5) indicates that degassing during the rise of the groundwater in the borehole and sampling was limited. For well 4, the estimated $CO_2$ degassing is 16%, whereas the $^{20}Ne/^{22}Ne$ ratio suggests that 67% (n=1) to 81% (n=2/3) of Ne was lost. An early degassing instead of a late degassing is further supported by an increasing degree of degassing with increasing brine proportion, which suggests that the brine had already degassed prior to mixing with the other groundwater components (see Fig. 5 for a modified mixing model).

An early degassing of the brine component before the mixing is difficult to quantify today, as most of the atmospheric noble gases in the samples today will derive from the other two mixing components, even for relatively high brine fractions. A late degassing, in turn, can be estimated and corrected for using Equations 3 and 4. The calculated proportion $f$ of Ne that remained in the groundwater samples decreases from 95-97% (for n=2/3 to 1) to 19-34% with increasing salinity. Noble gas concentrations $C_i^0$ corrected for a late degassing for n=1 are higher and show better agreement with the mixing model for an evaporative origin of the brine for the more saline samples, but the light noble gases like Ne are corrected too much relative to the heavy noble gases like Xe. Using n=2/3 for the degassing correction results in a more balanced correction of light and heavy noble gases (Fig. 5). For the most saline samples, corrected concentrations are above the mixing model for an evaporative brine origin in Fig. 5 and could be explained similarly well by membrane filtration. However, these corrected values are an upper limit, as the $^{20}Ne/^{22}Ne$ isotopic ratio could also be lowered by accumulation of nucleogenically produced Ne with a lower $^{20}Ne/^{22}Ne$ ratio, discussed in detail in section 5.3.2. Assuming the brine component before mixing had a $^{20}Ne/^{22}Ne$ ratio of 9.65 instead of 9.80 (black mixing line in Fig. 5e) due to nucleogenic production is completely consistent with the $^{20}Ne/^{22}Ne$ ratios measured in wells 1, 2, 5, and 7*. For wells 3 and 4, additional diffusive degassing is still required (59% and 64% of Ne lost, respectively). Degassing corrected concentrations in this case are entirely consistent with an evaporative origin of the brine (Fig. 5a-d).

It is worthwhile to note that neither incomplete gas extraction during sampling nor timing or mechanism of the degassing affect $^{81}Kr$ and $^{39}Ar$ dating or $^{85}Kr$ significantly because all these methods are based on isotope ratios with small mass differences (Purtschert et al., 2013). This is in contrast to accumulation ages deduced from radiogenic $^{4}He^*$ and $^{40}Ar^*$, for which timing and magnitude of the degassing are very critical.

## 5.2. Mixing Proportions

Based on the conclusions so far, we adopt and extend the mixing model proposed by Raidla et al. (2009) to the deeper parts of the CAS (Fig. 6). In their model, the brine component was assumed to be Paleozoic seawater with a $Cl^-$ content of 25 g/L and a $\delta^{18}O$ value of –2‰ VSMOW. In the present study, the highest $Cl^-$ content of groundwater is 90 g/kg (well 6, no noble gas data), which was chosen as an adjusted brine end member composition. The oxygen isotope composition in turn was shifted to 0‰ VSMOW to be more consistent with the evaporative genesis of the brine (Fig. 6). It can be shown that more enriched or depleted $\delta^{18}O$ values (e.g. ±2‰ VSMOW), albeit slightly altering the mixing proportions, would not significantly affect the interpretation of the noble gas concentrations and of the dating tracers in the next section. As a further simplification, mixing is assumed to be a non-fractionating, chemically conservative process. As shown in Fig. 6, the subspace of the modified mixing model indeed delineates the observed $Cl^-$ and $\delta^{18}O$ values of the deep groundwaters from the CAS, except for well 6, and is consistent with the proposed origin of the brine from evaporated seawater. In terms of groundwater residence times, an evaporated-seawater origin of the brine end member constrains the formation time of this end member to certainly pre-Quaternary and more likely even much earlier times. For the glacial meltwater end member in turn, a Pleistocene origin is plausible. The third component was considered to be of Holocene origin in Northern Estonia (Raidla et al., 2009) and consequently was named “recent”. However, in the deeper parts of the CAS, this third component could be considerably older, originating from previous interglacial periods. In this paper, it is thus called simply “meteoric” component, even though the glacial component is also of a meteoric origin.

Mixing proportions $f_i^{H2O}$ of the three end members calculated from $Cl^-$ and $\delta^{18}O$ for each groundwater sample from the CAS are given in Table 6. As an extension to here presented data, end-member proportions were also calculated for selected samples from Northern Estonia (NE1–NE3, cf. Fig. 6), using the $\delta^{18}O$ and $Cl^-$ values reported by Raidla et al. (2009). More details on the calculation of the mixing proportions are provided in the Electronic Appendix B. Four of the seven deep groundwater samples are dominated by the brine component (>50%) and in three of the samples (5–7*), the glacial meltwater component is absent or insignificant (Table 6). However, for the most saline samples that plot close to the brine end member, the distinction between the dilution of the brine by meteoric water and/or glacial meltwater can no longer be adequately resolved due to the uncertainty in the exact signature of the brine end member. Fig. 7 shows that glacial meltwater intruded the CAS mostly in the northern part of the BAB (Estonia), i.e. north of the fault zone, and did not reach Lithuania. In turn, the brine proportion is highest in wells in the deeper central part of the BAB and lower towards the boundaries of the BAB where the CAS is closer to the surface.

The mixing proportions $f_i^{Kr}$ of krypton, which are needed for the interpretation of the $^{81}Kr$ data, are then calculated using Equation 7, based on the $f_i^{H2O}$ and the Kr concentrations of the three end members. Noble gas concentrations of the end members were estimated in the following way: For the meteoric water, an infiltration temperature of 5°C, an elevation of the recharge area of 200 masl, unfractionated excess air corresponding to ΔNe=25%, and negligible salinity were assumed. The noble gas concentrations for the glacial end member are based on the median of concentrations measured in samples from Northern Estonia with a clear glacial meltwater signature in the isotope composition (Weißbach, 2014, for numerical values see Electronic Appendix B). The noble gas signature of the brine end member is chosen as AEW at a salinity of 163 g/kg (corresponding to 90 g/kg of $Cl^-$ for seawater-like composition) and a temperature of 25°C. For Kr, the respective concentrations are $11\cdot10^{-8}$, $19\cdot10^{-8}$, and $2\cdot10^{-8}$ $cm^3STP/g$ for meteoric water, glacial meltwater, and brine. The resulting noble gas concentration-$Cl^-$ mixing subspace, as denoted by the mixing lines in Fig. 5, is consistent with the measured pattern, with the exception of wells 3 and 4, for which a noble gas deficit is observed. This deficit is attributed to degassing of the groundwater. For a degassing during the genesis of the brine, this would lead to a lower noble gas content of the brine end member and consequently shifted mixing lines (Fig. 5). For a degassing during the rise of the groundwater in the borehole, the correction for diffusive degassing in Equations 3 and 4 bring the wells 3 and 4 in better alignment with the unmodified mixing lines. A direct comparison of modelled Kr concentrations with measured concentrations is provided in Electronic Appendix B.

## 5.3. Groundwater Residence Times

### 5.3.1. $^{81}$Kr Ages (of the glacial and meteoric water)

In the CAS, (apparent) $^{81}$Kr ages range from 319 ka to more than 1.3 Ma (Table 7). The positive correlation between $^{81}$Kr ages and brine proportions implies that the brine is the oldest water component, that is, older than 1.3 Ma – outside the dating range of $^{81}$Kr. This is the first time that $^{81}$Kr groundwater ages beyond the $^{81}$Kr dating range are reported. For the deconvolution of ages, it can thus be assumed that $R_{brine}/R_{air} < 0.02 \approx 0$, which reduces the number of unknowns in Equation 7 by one. For samples not containing any glacial meltwater, the meteoric component can thus be dated.

To deconvolute the $^{81}$Kr ages, it is very important to consider the Kr proportions $f_i'$ (see Equation 7), which are quite different from the water proportions (Fig. 7). Due to the low noble gas concentrations in the brine, the meteoric and glacial components dominate the budget of noble gas isotopic tracers like $^{81}$Kr even for relatively high brine water proportions. Therefore, the $^{81}$Kr age mainly represents the age of the meteoric water and glacial meltwater components, except for samples with very high brine proportions. This distinction between water proportions and noble gas proportions is often not made when analyzing isotopic tracers, but is crucial for a correct interpretation of $^{81}$Kr in this study.

Residence times of the meteoric and glacial components are unlikely to be constant throughout the whole CAS, due to the highly variable thickness of the overburden, the large spatial extent and a predominately horizontal flow (at least today). However, well-specific $R_{glacial}$ and $R_{meteoric}$ mean that Equation 7 is underdetermined when only using $^{81}$Kr. Thus, for the wells containing glacial meltwater, especially wells 1 and 2, further assumptions are required. For example, it could be assumed that all the glacial meltwater originates from the last deglaciation and is therefore between 10 and 25 ka old. However, this would result in negative $^{81}$Kr abundances of the meteoric component $R_{meteoric}$, which is physically impossible. Therefore, the glacial component must be a mixture of meltwater from at least the last two glaciations implying that water from (the) previous glacial period(s) was not completely flushed from the CAS during the last interglacial period. To solve Equation 7, it is thus assumed that the glacial meltwater and the meteoric water in any one well have approximately the same age, resulting in $^{81}Kr^{m}$ ages of the meteoric and glacial components that are similar to the $^{81}$Kr ages, ranging from 300 ka to over 1.3 Ma (Table 7). Because only a lower age limit of 1.3 Ma can be given for the brine, only a lower limit can be given for $t_{mix}$ (Equation 6), which ranges from 400 ka to 1.3 Ma (Table 7).

### 5.3.2. He, Ne, and Ar Accumulation Ages (of the brine)

The very old groundwater ages deduced from $^{81}Kr$ are also supported by the elevated $^{4}He$ concentrations and $^{40}Ar/^{36}Ar$ ratios in samples 3, 4, and 7 (cf. Table 3, Torgersen and Stute, 2013). Especially such elevated $^{40}Ar/^{36}Ar$ ratios are typically only observed in very old water with residence times of at least several hundred ka due to accumulation of radiogenic $^{40}Ar$. The radiogenic component, $^{40}Ar^*$, can be calculated by (Aeschbach-Hertig and Solomon, 2012):

$$^{40}Ar^* = {}^{40}Ar_{measured} - {}^{36}Ar_{measured} \cdot R_{atm}^{Ar} \tag{9}$$

For He, the atmospheric component is negligible compared to the radiogenic component He*.

$$^{4}He^* \approx {}^{4}He_{measured} \quad \text{and} \quad {}^{3}He^* \approx {}^{3}He_{measured} \tag{10}$$

Nucleogenic production of Ne also leads to an accumulation of both $^{20}Ne$ and $^{22}Ne$ (Kennedy et al., 1990, Ballentine and Burnard, 2002). The amount of $^{20}Ne^*$ can be estimated by:

$$^{20}Ne^* = \frac{1}{1 - R_{atm}^{Ne} \cdot \frac{p^{22}}{p^{20}}} \cdot \left({}^{20}Ne_{measured} - R_{atm}^{Ne} \cdot {}^{22}Ne_{measured}\right) \tag{11}$$

where $R_{atm}^{Ne}$ is the atmospheric $^{20}Ne/^{22}Ne$ ratio of 9.80 and $p^{22}/p^{20}$ is the nucleogenic production ratio of the two isotopes. All three equations neglect gas loss and isotopic fractionation in the subsurface e.g. during degassing. For a late degassing, measured values first have to be corrected for degassing using Equations 3 and 4. An early degassing would have a negligible effect on $^{4}He^*$ and $^{40}Ar^*$ estimations, but lead to a considerable overestimation of $^{20}Ne^*$ when using Equation 11 as the degassing would lower the initial $^{20}Ne/^{22}Ne$ ratio.

Crustal $^{4}He^*/^{40}Ar^*$ production ratios in the rocks of ~3.4 were calculated according to Ballentine and Burnard (2002) with K, U, and Th concentrations for the CAS from the literature: Raidla et al. (2006) ($K_2O_{eq}$ = 4% in Estonia) and Shogenova et al. (2003) (3% K, 1.5 ppm U and 7 ppm Th in three samples from Lithuania). All water samples show higher $^{4}He^*/^{40}Ar^*$ ratios than the production ratio in the rock (up to 4.5 times the production ratio, see Fig. 8a). This is attributed to preferential release of the smaller and lighter $^{4}He$ from the host minerals (Mamyrin and Tolstikhin, 1984; Ballentine et al., 1994). The higher the temperature, the higher the likelihood that the produced atoms are released into the water-filled pore space. The closure temperature above which most atoms are released is between 200 and 300°C for $^{40}Ar$ and ~100°C for $^{4}He$ (Ballentine and Burnard, 2002). Depending on the in-situ temperature and on the grain size distribution of the U and K bearing minerals, considerably higher $^{4}He^{*}/^{40}Ar^{*}$ ratios are possible.

For an estimation of $^{20}Ne^{*}$, the production ratio of $^{20}Ne^{*}/^{22}Ne^{*}$ is required, which is very sensitive to the O/F ratio of the rocks contributing the Ne. The average crustal value is O/F=752 (Ballentine and Burnard, 2002), but Kennedy et al. (1990) suggested that O/F=113 is more appropriate for the mineral environment in which the U and Th are typically sited. Based on Ballentine and Burnard (2002) and using K, U, and Th as given above, a Ne production ratio $p^{22}/p^{20}$ =19.53 is obtained for O/F=113 and the estimated $^{20}Ne^{*}$ ranges from 1.0 to $5.5 \cdot 10^{-12}$ cc/g. For O/F=752, the production ratio is 2.942, which also results in ~7 times higher $^{20}Ne^{*}$. Comparing the estimated $^{20}Ne^{*}$ to $^{4}He^{*}$ excludes O/F=752 as some samples already contain more $^{20}Ne^{*}$ than expected with O/F=113 (assuming 100% of produced Ne is released, see Fig. 8b). In fact, to explain the apparent $^{4}He^{*}/^{20}Ne^{*}$ ratio of sample 4, the O/F ratio would have to be as high as 45. In other words, either the O/F ratio or the released fraction of Ne have to vary by a factor of over 2.5 in the CAS.

Alternatively, degassing, which has not been taken into account so far, might be responsible for the differences in estimated $^{4}He^{*}/^{20}Ne^{*}$ ratios. Solubility-controlled degassing has only a relatively small effect on the $^{4}He^{*}/^{20}Ne^{*}$ ratio, even if the degassing happened after the accumulation (Fig. 8b) and for $^{20}Ne^{*}/^{40}Ar^{*}$ the effect is actually in the wrong direction (Fig. 8c). An early diffusive degassing (before the accumulation started) does not affect $^{4}He^{*}$ and $^{40}Ar^{*}$significantly but would result in a reduced initial $^{20}Ne/^{22}Ne$ ratio. This results in an overestimation of the produced $^{20}Ne^{*}$ when using Equation 11. The initial $^{20}Ne/^{22}Ne$ ratio can be calculated from the $^{20}Ne^{*}$ value obtained when tracing the red line (parallel to the production line) in Fig. 8b to $^{4}He^{*}$=0. The "apparent" initial $^{20}Ne^{*}$ is 3.6E-12 cc/g for wells 3 and 4, corresponding to an initial $^{20}Ne/^{22}Ne$ ratio of 9.75 instead of 9.80. The effect of late degassing (during the ascent in the well) is also shown in Fig. 8. The underlying assumption is that, at first, radiogenic and nucleogenic production increase $^{4}He^{*}$, $^{20}Ne^{*}$, and $^{40}Ar^{*}$ along the (release-modified) in-situ production lines. Then, diffusion- or solubility-controlled degassing according to Equations 3 and 4 takes place and finally, $^{3}He^{*}$, $^{20}Ne^{*}$, and $^{40}Ar^{*}$ are calculated according to Equations 9-11. Interestingly, during the degassing, the "apparent" $^{20}Ne^{*}$ increases initially, before it starts decreasing. The reason is that the degassing reduces not only the Ne concentration in the groundwater, resulting in less "apparent" $^{20}Ne^{*}$, but at the same time also reduces the $^{20}Ne/^{22}Ne$ ratio, resulting in more "apparent" $^{20}Ne^{*}$. The first effect dominates if more than 50% of the Ne is lost, whereas the second effect dominates initially, until 50% of the Ne is lost. Fig. 8b shows that samples 3 and 4 lost ~55% and ~60% of the initial Ne by diffusive degassing, respectively. Fig. 8a gives a similar value for well 4, but shows no significant degassing for well 3. However, the result from Fig. 8b is more robust, as it is independent of the only weakly constrained release fraction of $^{40}$Ar*. The degrees degassing estimated from Fig. 8b are very close to those estimated based on the $^{20}$Ne/$^{22}$Ne ratios as a function of salinity in section 5.1.3 and Fig. 5e. They thus also lead to very similar corrections for the noble gas concentrations in Fig. 5a-d, consistent with an evaporative origin of the brine. In summary, the measured $^{20}$Ne/$^{22}$Ne ratios of wells 3 and 4 are most consistent with a combination of diffusive degassing and nucleogenic production with O/F=113. Early and late degassing both seem possible, but they differ considerably in the amount of $^{4}$He*, $^{20}$Ne*, and $^{40}$Ar* accumulated in samples 3 and 4. Unfortunately, concentrations of $^{21}$Ne and $^{38}$Ar, which would provide much better constraints on the relative importance of nucleogenic production and mass fractionation, were not measured.

The concentrations of $^{4}$He*, $^{20}$Ne*, and $^{40}$Ar* are often used as qualitative dating tools because their absolute accumulation rates *r* are difficult to determine. Qualitatively, the $^{4}$He*, $^{20}$Ne*, and $^{40}$Ar* systematics agree with the conclusions drawn from $^{81}$Kr activities: The three wells with higher $^{81}$Kr ages (wells 3, 4, and 7*) also exhibit higher $^{4}$He* and $^{40}$Ar* concentrations (see Table 3 and 7). In several studies, the $^{4}$He* accumulation rate was calibrated using $^{14}$C (Torgersen and Clarke, 1985; Marty et al., 1993) or $^{81}$Kr (Lehmann et al., 2003; Purtschert et al., 2013; Aggarwal et al., 2015). However, this approach is problematic here, because $^{81}$Kr mainly dates the meteoric and glacial component, whereas a large proportion of the $^{4}$He*, $^{20}$Ne*, and $^{40}$Ar* originates from the older brine component. Therefore, $^{4}$He*, $^{20}$Ne*, and $^{40}$Ar* concentrations are related to the proportion of brine water ($f_{brine}$), as shown in Fig. 9, rather than to the groundwater ages deduced from $^{81}$Kr. For wells 3 and 4, uncorrected values, values corrected for an early degassing and values corrected for a late degassing are shown. The expected linear relationships of $^{4}$He*, $^{20}$Ne*, and $^{40}$Ar* with the brine fraction fit best with the late-degassing values, providing further evidence for a late degassing.

Ranges of $^{4}$He*, $^{20}$Ne*, and $^{40}$Ar* concentrations of the meteoric/glacial component and the brine end member are estimated by extrapolating the linear fit in Fig. 9 to a brine proportion of 0 and 1, respectively: $<1.5\cdot10^{-4}$ cm$^3$ (STP)/$g_{water}$ and $1.72\cdot10^{-3}$ cm$^3$ (STP)/$g_{water}$ for $^{4}$He*, $<1.5\cdot10^{-12}$ cm$^3$ (STP)/$g_{water}$ and $6.8\cdot10^{-12}$ cm$^3$ (STP)/$g_{water}$ for $^{20}$Ne*, and $<2.0\cdot10^{-5}$ cm$^3$ (STP)/$g_{water}$ and $9.1\cdot10^{-5}$ cm$^3$ (STP)/$g_{water}$ for $^{40}$Ar*. The deduced ratios of $^{4}$He*/$^{20}$Ne*, $^{4}$He*/$^{40}$Ar*, and $^{20}$Ne*/$^{40}$Ar* of the brine component are close to the respective ratios of in-situ production rates.

A calibration of $^{4}$He*, $^{20}$Ne*, and $^{40}$Ar* accumulation rates by means of $^{81}$Kr data is limited to samples where the brine proportion is small (wells 1 and 2), because groundwater ages deduced from $^{81}$Kr do not capture the brine age. The $^{81}$Kr age scale shown Fig. 9 is based on the mixed $^{81}$Kr ages ($t_{mix}$, Equation 6) of samples 1 and 2 (Table 7). These are minimal age estimates since the brine component might be older than 1.3 Ma. For comparison, age scales corresponding to in-situ production and an external influx (Torgersen and Clarke, 1985) are also shown. In-situ production rates (assuming 100% release) were calculated according to Ballentine and Burnard (2002) for a rock density of 2.7 g/cm$^3$ and a porosity of 5% (porosity of the shallow parts of the CAS is 20% or more, but has been reduced to less than 5% in 2000 m depth by sediment compaction (Raidla et al., 2006)). Accumulation rates for an external influx were estimated based on crustal fluxes reported by Torgersen (2010) for $^{4}$He* and Torgersen et al. (1989) for $^{40}$Ar* and assuming that the top of the CAS is impermeable (i.e. no gas is lost to overlying strata). The $^{20}$Ne* flux was calculated based on the $^{4}$He* flux and the $^{20}$Ne*/$^{4}$He* production ratio. On the $^{81}$Kr-calibrated age scale, $^{4}$He*, $^{20}$Ne*, and $^{40}$Ar* ages of the meteoric/glacial component are <450 ka, <650 ka and <1.0 Ma, respectively. The $^{81}$Kr-calibrated $^{4}$He*, $^{20}$Ne*, and $^{40}$Ar* ages of the brine component in turn are 5.0±0.8 Ma, 3.0±0.6 Ma, and 4.5±1.3 Ma, respectively. Overall, $^{4}$He* and $^{40}$Ar* qualitatively support the groundwater ages deduced from $^{81}$Kr and suggest that the brine is pre-Quaternary, in line with the evaporative origin of the brine proposed in section 5.1.1.

## 5.4. Recharge Dynamics and Flow Pattern

### 5.4.1. Spatial Pattern of $^{81}$Kr Ages and He and Ar Accumulation Ages

The $^{81}$Kr$^{m}$ ages of the meteoric and glacial water show a higher correlation with depth below surface (Fig. 10b) than with distance from the present day recharge area in the southeast (Fig. 10a). This is the combined result of (i) the concave shape of the basin, translating horizontal movement into a vertical stratification and (ii) the present-day distribution of $^{81}$Kr ages reflecting net groundwater flow direction and velocity over the last 1 Ma, integrating over several glacial cycles with complex changes of the hydraulic regime. The correlation with depth does not mean, however, that the groundwater is stagnant, as diffusion would be too slow to transport enough $^{81}$Kr downwards.

A similar spatial pattern as for $^{81}$Kr$^{m}$ is also observed for $^{4}$He* and $^{40}$Ar*. The observed patterns can be explained neither by the flow pattern implied by the present-day hydraulic heads nor by assuming that the water is essentially stagnant on time scales of several hundred thousands of years. This implies that either flow patterns have been different in the past, possibly due to hydraulic reorganization during glaciations (Person et al., 2007), or that vertical exchange between the CAS and overlying formations is much larger at least in some areas, resulting in a more complex present-day flow pattern than the one presented so far.

An aquifer as large as the CAS will take a considerable time to adjust to the change in hydraulic (head) conditions. Rousseau-Gueutin et al. (2013) proposed a method to estimate the time $t_{NE}$ needed for an aquifer to reach a near-steady state (NE = near-equilibrium) following a large hydraulic perturbation. For a confined aquifer with horizontal flow, it is

$$t_{NE} = 3 \cdot \frac{4 S_S L^2}{K_h \pi^2} \quad (12)$$

$t_{NE}$ is thereby defined as the time after an initial perturbation dissipated by an average of 95% across the aquifer. $L$ is the aquifer length (650 km), $K_h$ the hydraulic conductivity (2.95 m/day in the CAS), and $S_s$ the specific storage. According to Domenico and Mifflin (1965), the specific storage of material from dense sand to fissured rock is on the order of $10^{-4}$ $m^{-1}$. These numbers result in a relaxation time of $t_{Ne}$ ~50 ka suggesting that the CAS is probably still recovering from the last glaciation. With two changes of hydraulic conditions during each full glacial cycle of ~100 ka, the CAS was in transient states for most of the time over the last 1 Ma. Long relaxation times act as a low pass filter, dampening and delaying the system's response to changing hydraulic conditions.

### 5.4.2. Transient Groundwater Flow Over Several Glacial Cycles

The purpose of this section is to present a simple conceptual model of groundwater flow over several glacial cycles that is in accordance with the observed tracer data and its spatial patterns. The simplified conceptual model assumes that the Ordovician-Silurian (O-S) aquiclude is impermeable, isolating the CAS completely except in the recharge and discharge zones in the southeast and the Baltic Sea. In light of the complexity and size of the CAS and the small number of $^{81}$Kr samples, it would not be sound to draw detailed or quantitative conclusions from the conceptual model. However, we believe that the conceptual model is valuable in recommending relevant future research directions to better understand long-term groundwater flow in the CAS.

Hydrogeochemical studies (Vaikmäe et al., 2008; Raidla et al., 2009), noble gas studies (Vaikmäe et al., 2001), and hydrodynamic modeling (Saks et al., 2012) suggest the intrusion of glacial meltwater – most likely in the outcrop area (Raidla et al., 2012; Saks et al., 2012) and probably through buried valleys in Northern Estonia (Vaikmäe et al., 2001) – and a regional reversal of the flow direction during the last glaciation. Such an intrusion may be the reason for the relatively young $^{81}$Kr$^{m}$ ages of the glacial and meteoric component in wells 1 and 2. Studies in other regions across the northern hemisphere found that glacial water may reach depths of up to 1000 m and smaller aquifers can be flushed in a few ka because of the significantly higher hydraulic gradients during the LGM (Boulton et al., 1996; Piotrowski, 1997; Lemieux et al., 2008; Lemieux and Sudicky, 2010; McIntosh et al., 2012). However, due to the large size of the BAB, one glacial cycle is insufficient to flush the whole system (Saks et al., 2012). Over several glacial cycles, multiple reversals of the flow direction in the CAS occurred (blue arrows in Fig. 10a).

Conceptually, throughout a glacial cycle, four stages of glacial overburden and groundwater flow patterns can be distinguished in the CAS (Fig. 11). Stage 1 (interglacial) represents the present-day situation with recharge in the southeast and discharge in the northwest of the BAB. Stage 2 prevails at the beginning and end of the glaciation, when the present-day discharge area in the north is covered by ice, but the fault zone is still ice-free. In case of vertical leakage across the fault zone, it now acts as a discharge zone because of the highly elevated hydraulic head underneath the ice sheet. In case of no leakage, the flow pattern is identical to stage 3, which is reached when the fault zone is also ice covered. At this point, the aquifer system discharges to the present-day recharge area, whereas recharge takes place below the glacier. The increasing lithostatic pressure from the glacier might reduce the conductivity of the fractures in the fault zone, reducing vertical leakage or blocking it completely (Saks et al., 2012). In Stage 4, the whole BAB is covered by ice and groundwater flow is driven by the hydrostatic pressure at the ice sheet base, which follows ice sheet topography. Stage 4 was probably not reached during the last glaciation (Weichselian), but was so during the Saalian (Guobyte and Satkunas, 2011). During deglaciation, the stages are passed through in reverse order. For simplicity, the effects of permafrost on groundwater flow are not discussed in detail here (see for example Boulton et al., 1996; Lemieux et al., 2008; McIntosh et al., 2012; for the BAB: Jõeleht, 1998; Zuzevicius, 2010; Saks et al., 2012). The interrelations between permafrost, glaciers and groundwater flow are complex, but will likely not significantly change the conceptual model presented above, other than some stages lasting for a slightly longer or shorter time.

The net flow rate of this conceptual model over the last 1 Ma depends on the balance between volumetric fluxes in the present-day flow direction and in the reverse direction integrated over a full glacial cycle. A reconstruction of the ice sheet extent during the last glacial cycle based on lithostratigraphical, biostratigraphical, and geochronological information (Guobyte and Satkunas, 2011) provides evidence that stages 2 and 3 were prevalent for considerable periods during the last glacial cycle. The temperature-corrected global marine $\delta^{18}O$ record, which is a proxy for global ice volume, indicates that over the last 1 Ma, several glaciations occurred with similar or even larger ice volumes (Shakun et al., 2015). Furthermore, a shorter duration of the stages with reverse flow direction might be compensated for by higher volumetric fluxes during these stages due to the larger hydraulic gradients during the reversed flow (Jõeleht, 1998). Finally, if the integrated volumetric flux in one of the two directions dominates markedly, a continuous increase of groundwater ages in the net flow direction results, which is in conflict with the observed pattern of $^{81}Kr^{m}$ ages. With integrated volumetric fluxes of a similar magnitude in both directions, a back-and-forth movement results. The resulting integrated net flux and net seepage velocity in the direction with the larger volumetric flux is thus smaller than the average modulus of the instantaneous volumetric flux rate and seepage velocity. This would explain why our $^{81}Kr^{m}$ ages suggest a smaller seepage velocity than the one deduced by Virbulis et al. (2013, Fig. 10c). Even for a zero net flux, there is still a $^{81}Kr$ concentration gradient and transport of $^{81}$Kr due to dispersion. A rough estimate shows that a dispersivity of ~10km could potentially reproduce the $^{81}$Kr$^{m}$ ages (Fig. 10c.).

So far, it was assumed that the CAS only gains or loses water in the current recharge and discharge areas only. However, topography driven recharge and discharge fluxes on sub-basin scales (Tóth, 1963), and especially vertical exchange across the Liepāja-Pskov fault zone might also occur. The reduced net flux resulting from the back-and-forth movement of our conceptual model indicates that even relatively small leakage fluxes might be relevant for the long-term water budget (purple arrows in Fig. 10a).

Overall, there is some evidence from our data that flow patterns in the past differed significantly from the present-day pattern, with reversed flow during glaciations. Furthermore, the system must have been in a transient state for much of the last 1 Ma. Finally, understanding flow in the Liepāja-Pskov fault zone and other (even relatively small) leakage fluxes is crucial for an improved understanding of the flow pattern in the CAS as a whole. Apart from further modelling attempts, sampling additional wells for $^{81}$Kr dating along a flow line from the Belarus-Masurian Anticline to the Baltic Sea and from the fault zone towards the Moscow Basin would help to test our hypotheses.

## 6. Conclusions

In this study of the Cambrian Aquifer System in the Baltic Artesian Basin, chemistry, stable isotopes, noble gases, and dating tracers were combined for a better understanding of flow and recharge dynamics of the system over the last one million years. We find that the variability in chemical composition, stable isotopes and noble gas concentrations in the basin is predominately controlled by mixing of three distinct water masses: Holocene and Pleistocene interglacial meteoric water, glacial meltwater, and a brine end member. Thanks to its inertness, constant atmospheric input function, and its insensitivity to degassing, $^{81}$Kr turns out to be a nearly ideal dating tracer for such old waters. This is the first groundwater study with $^{81}$Kr activities below the detection limit of the ATTA-3 instrument, currently at 2% of the atmospheric $^{81}$Kr/Kr ratio. Our results confirm that under normal conditions, underground production of $^{81}$Kr is not affecting the $^{81}$Kr dating results. However, the differing noble gas concentrations of the different water components, in particular the depleted noble gas concentrations of the brine, have to be considered when interpreting measured $^{81}$Kr activities in terms of groundwater age. Diffusive loss of $^{81}$Kr to stagnant water is another process that might affect groundwater ages deduced from $^{81}$Kr (Purtschert et al., 2013; Sturchio et al., 2014) but was neglected in this study because of the relatively high porosity and large thickness of the CAS. The radiogenic $^{4}$He* and $^{40}$Ar* concentrations provide additional age information, but are more difficult to interpret. Qualitatively they support and complement the age structure derived from the $^{81}$Kr data as they “date” mainly the brine component, which is beyond the dating range of $^{81}$Kr.

The dating tracers $^{81}$Kr, $^{4}$He*, and $^{40}$Ar* indicate a residence time of the brine component of more than 1–5 Ma. Some uncertainty about the brine formation process remains, but the combination of chemical and stable isotope composition of the brine, noble gas concentrations and dating results favors evaporative enrichment of seawater, implying a pre-Quaternary origin of the brine. Tracer ages of meteoric water and glacial meltwater are on the order of several hundred thousand years, so that the possibility of multiple reversals of the flow direction in the Cambrian Aquifer System as a result of the paleoclimatology of the area has to be taken into account. Under such conditions, small vertical leakage, through fracture zones for example, might have a considerable impact on the net flow pattern. Due to the cyclic changes in hydraulic conditions and potential reversals of flow direction, the aquifer was probably in a transient state over most of the last 1 Ma period, which needs to be considered in future modeling attempts.

The conclusions we can draw are limited by the small number of samples from the deeper parts of the aquifer system. Furthermore, an improved modelling approach concerning the faults and their effect on flow dynamics could help to test different potential explanations of the spatial age pattern found in this study. More research is also needed on the role of permafrost in blocking recharge or discharge pathways. Finally, noble gas measurements from Estonia, where the proportion of glacial meltwater is high, noble gas derived infiltration temperatures and stable isotopes could be used to differentiate the glacial recharge mechanism (basal melting or surficial meltwater) as Grundl et al. (2013) have done for groundwater recharged from the Laurentide Ice Sheet in southeastern Wisconsin.

## Acknowledgements

Sampling would not have been possible without the kind help from Värska Spa and Goodman’s (Häädemeeste) in south Estonia. Valuable assistance in the field was given by Jüri Ivask from the Institute of Geology at Tallinn University of Technology. Furthermore, we thank Peter Nyfeler, who performed the stable isotope analyses in Bern and Lauren Raghoo, who provided helpful feedback on the manuscript. W.J., Z-T.L., P.M., J.C.Z. and the Laboratory for Radiokrypton Dating at Argonne are supported by DOE, Office of Nuclear Physics, under contract DE-AC02-06CH11357. Development of the ATTA-3 instrument was supported in part by NSF EAR-0651161. This study was supported by the Estonian Research Council (grant IUT19-22 to RV and PUTJD127 to VR). The study is a contribution to the INQUA and UNESCO supported G@GPS Project.

Table 1: Depth and location of the sampled wells in the Cambrian and Devonian (only 7*) aquifer systems.

| ID | Well number | Well name | Aquifer | Country | Coordinates | | Depth |
|---|---|---|---|---|---|---|---|
| | | | | | (°N) | (°E) | (m) |
| 4613 | **1** | **Värska** | Cm-V | Estonia | 57.99 | 27.63 | 600 |
| 8021 | **2** | **Häädemeeste** | Cm | Estonia | 58.08 | 24.50 | 610 |
| 50194 | **3** | **Riga** | Cm | Latvia | 56.97 | 24.25 | 1027 |
| 50202 | **4** | **Kemeri** | Cm | Latvia | 56.97 | 23.56 | 999 |
| 50423 | **5** | **Ignalina** | Cm | Lithuania | 55.34 | 26.15 | 500 |
| 14R2 | **6** | **Genciai** | Cm | Lithuania | 55.87 | 21.18 | 1800 |
| 25872 | **7*** | **Klaipeda** | $D_1$ | Lithuania | 55.68 | 21.20 | 1100 |

Table 2: Field measurements and chemical and stable isotopic composition of groundwater from the deep wells in the BAB. Samples were collected together with those for $^{81}Kr$ analyses during campaign II (see section 3). The density was calculated from temperature and salinity (Millero and Huang, 2009). In all samples, concentrations of $NH_4^+$ and $NO_3^-$ were below their respective detection limits of <10mg/kg and <1.6 mg/kg.

| Well | Sampling Date | Density (g/cm$^3$) | pH | Eh (mV) | Temperature[a] (°C) | $Na^+$ | $K^+$ | $Li^+$ | $Ca^{2+}$ | $Mg^{2+}$ | $Sr^{2+}$ | $Ba^{2+}$ | $Mn_{tot}$ (mg/kg) | $Cl^-$ | $Br^-$ | $SO_4^{2-}$ | TIC | TOC | Si | B | TDS | Charge Balance | $\delta^{18}O$ | $\delta^2H$ (‰ VSMOW) | D-Excess |
|---|---|---|---|---|---|---|---|---|---|---|---|---|---|---|---|---|---|---|---|---|---|---|---|---|---|
| **1** | 01-Oct-12 | 1.0118 | 7.22 | -21 | n.m. | 5286 | 79 | 0.55 | 994 | 379 | 21.4 | 0.21 | 1.04 | 11109 | 53 | 246 | 36.8 | 17.2 | 6.7 | 4.7 | 18182 | -1.2% | -12.63±0.02 | -92.8±0.3 | 8.21 |
| **2** | 02-Oct-12 | 1.0021 | 7.52 | -38 | 12.2 | 1645 | 36 | 0.28 | 175 | 78 | 3.9 | 0.18 | 0.06 | 3087 | 15 | 73 | 53.3 | 8.2 | 6.5 | 4.2 | 5127 | -2.4% | -13.61±0.04 | -100.7±0.1 | 8.14 |
| **3** | 03-Oct-12 | 1.0795 | 8.10[b] | -72 | 10.1 | 30580 | 317 | 2.40 | 6344 | 2378 | 129.5 | 0.20 | 6.74 | 64632 | 266 | 1277 | 5.0 | 12.0 | 4.6 | 5.9 | 105944 | 0.1% | -4.79±0.06 | -42.7±0.3 | -4.39 |
| **4** | 03-Oct-12 | 1.0784 | 7.07 | -12 | 9.7 | 29907 | 321 | 2.68 | 6434 | 2407 | 125.1 | 0.22 | 5.59 | 63996 | 268 | 1126 | 8.0 | 11.5 | 5.7 | 6.2 | 104605 | 0.0% | -4.44±0.17 | -40.6±0.3 | -5.04 |
| **5** | 04-Oct-12 | 1.0320 | 7.55 | -12 | 12.1 | 14084 | 184 | 0.98 | 1726 | 740 | 31.2 | 0.05 | 0.68 | 25531 | 79 | 2509 | 30.9 | 9.4 | 9.6 | 4.1 | 44901 | -0.6% | -7.23±0.04 | -55.2±0.3 | 2.69 |
| **6** | 05-Oct-12 | 1.1154 | n.m. | n.m. | n.m. | 29264 | 688 | 7.36 | 19347 | 3475 | 368.3 | 102.65 | 14.75 | 89664 | 741 | 15 | 5.1 | 43.5 | 4.8 | 18.7 | 143765 | 0.4% | -5.03±0.09 | -37.2±0.1 | 2.99 |
| **7*** | 05-Oct-12 | 1.0661 | 5.74[b] | -109.8 | 36.4 | 23400 | 557 | 3.31 | 6366 | 2175 | 142.8 | 0.16 | 0.86 | 53907 | 346 | 1606 | 16.6 | 12.0 | 6.6 | 11.0 | 88523 | -0.8% | -4.46±0.10 | -34.7±0.8 | 0.96 |

n.m. not measured
[a] measured at the surface. In-situ measurements in wells 6 and 7* were 35.5°C and 40°C, respectively
[b] unreliable field measurements

Table 3: Noble gas concentrations, isotopic ratios, and Ne excess of groundwater samples collected in campaign III (section 3).Concentrations of $N_2$ to $CH_4$ and the total dissolved gas content (TDG) are inferred from the gas extracted for $^{81}$Kr analysis in campaign II. For comparison, concentrations for air-equilibrated fresh water, seawater, and the brine end member are also given. Concentrations are given in cm$^3$STP/g if not indicated otherwise.

| Well | $^{3}$He ($10^{-12}$) | $^{4}$He ($10^{-4}$) | $^{20}$Ne ($10^{-8}$) | $^{22}$Ne ($10^{-9}$) | $^{36}$Ar ($10^{-7}$) | $^{40}$Ar ($10^{-5}$) | $^{84}$Kr ($10^{-8}$) | $^{132}$Xe ($10^{-10}$) | $^{20}$Ne/$^{22}$Ne | $^{40}$Ar/$^{36}$Ar | $^{4}$He/$^{20}$Ne | $^{3}$He/$^{4}$He ($10^{-8}$) | $\Delta Ne_s$[a] | $\Delta Ne_{mix}$[b] | $N_2$ ($10^{-2}$) | $O_2$ ($10^{-3}$) | Ar ($10^{-5}$) | $CO_2$ ($10^{-5}$) | $CH_4$ ($10^{-5}$) | TDG[c] (cm$^3$STP/L) |
|---|---|---|---|---|---|---|---|---|---|---|---|---|---|---|---|---|---|---|---|---|
| 1 | 3.4 ± 0.1 | 1.77 ± 0.01 | 25.25 ± 0.07 | 25.79 ± 0.07 | 15.8 ± 0.4 | 48.03 ± 0.37 | 5.88 ± 0.06 | 37.2 ± 0.7 | 9.79 ± 0.04 | 304 ± 7 | 700 ± 5 | 1.91 ± 0.07 | 51% | 99% | 2.35 | 0.03 | 43.9 | 5.3 | 7.5 | 28.3 |
| 2 | n.m. | 2.04 ± 0.02 | 37.46 ± 0.13 | 38.25 ± 0.15 | 19.8 ± 0.5 | 60.47 ± 0.09 | 7.05 ± 0.06 | 44.4 ± 0.8 | 9.80 ± 0.05 | 305 ± 7 | 544 ± 6 | n.m. | 112% | 93% | 2.45 | 0.01 | 48.1 | 4.1 | 0.5 | 29.4 |
| 3 | 9.2 ± 0.3 | 4.43 ± 0.02 | 7.18 ± 0.02 | 7.43 ± 0.02 | 5.4 ± 0.4 | 19.30 ± 0.04 | 2.06 ± 0.02 | 10.8 ± 0.2 | 9.66 ± 0.03 | 355 ± 26 | 6'174 ± 35 | 2.08 ± 0.08 | -25% | -30% | 1.58 | 0.01 | 14.8 | 0.2 | 1.6 | 18.7 |
| 4 | 5.6 ± 0.3 | 2.81 ± 0.02 | 5.66 ± 0.01 | 5.86 ± 0.01 | 3.5 ± 0.3 | 14.63 ± 0.04 | 1.74 ± 0.02 | 9.3 ± 0.2 | 9.65 ± 0.03 | 419 ± 39 | 4'961 ± 29 | 1.99 ± 0.10 | -41% | -44% | 2.37 | 0.02 | 19.3 | 0.9 | 10.1 | 28.2 |
| 5 | 3.0 ± 0.1 | 1.76 ± 0.01 | 21.63 ± 0.06 | 22.11 ± 0.06 | 13.3 ± 0.4 | 41.14 ± 0.07 | 5.14 ± 0.04 | 33.8 ± 0.6 | 9.79 ± 0.04 | 310 ± 10 | 812 ± 6 | 1.69 ± 0.06 | 54% | 36% | 1.79 | 0.02 | 31.5 | 2.4 | 3.2 | 21.6 |
| 6 | n.m. | n.m. | n.m. | n.m. | n.m. | n.m. | n.m. | n.m. | n.m. | n.m. | n.m. | n.m. | n.m. | n.m. | 3.63 | 0.05 | 22.4 | 13.0 | 5'458.7 | 107.4 |
| 7* | 19.2 ± 0.6 | 8.12 ± 0.05 | 14.52 ± 0.02 | 14.88 ± 0.04 | 9.5 ± 0.5 | 33.47 ± 0.06 | 3.59 ± 0.03 | 20.0 ± 0.3 | 9.76 ± 0.03 | 353 ± 18 | 5'594 ± 36 | 2.37 ± 0.08 | 42% | 23% | 3.14 | 0.04 | 30.3 | 20.5 | 0.1 | 37.5 |
| Fresh water[d] | 0.067 | 0.00048 | 19.6 | 20.0 | 14.8 | 44.1 | 6.1 | 43.1 | | | | | | | 1.62 | 7.3 | 44.1 | 46.9 | 0.007 | |
| Seawater[e] | 0.056 | 0.00040 | 15.7 | 16.0 | 11.4 | 34.1 | 4.7 | 32.2 | | | | | | | 1.23 | | 34.1 | | | |
| Brine end member[f] | 0.028 | 0.00020 | 6.6 | 6.7 | 3.2 | 9.6 | 1.1 | 6.6 | | | | | | | 0.45 | | 9.6 | | | |

[a] Calculated for an altitude of 0 m asl and temperature and TDS from Table 2 for each well. For well 1, T=10°C was assumed
[b] Calculated according to Equation 2. For the glacial and meteoric water, negligible salinity and a mean infiltration temperature of 0°C and 5°C, respectively, are assumed. For the brine, infiltration conditions as given in [f] are assumed.
[c] Assuming an extraction yield of 85%
[d] Calculated for an altitude of 0 m asl, a temperature of 5°C, and TDS=0 g/kg
[e] Calculated for an altitude of 0 m asl, a temperature of 5°C, and TDS = 35 g/kg
[f] Calculated for a temperature of 25°C, an altitude of 0 m asl and TDS = 163 g/kg (corresponding to 90 g/kg of Cl- for a chemical composition like seawater)
n.m. Not measured

Table 4: Measurements of the dating tracers and contamination (Cont) corrected results for $^{81}$Kr and $^{39}$Ar. Samples for $^{14}$C dating were collected in campaign I, samples for $^{85}$Kr, $^{81}$Kr, and $^{39}$Ar in campaign II (section 3).

| Well | **$^{85}$Kr** (dpm/cc$_{Kr}$) | **$^{39}$Ar** (%modern) | **$^{14}$C** (pmc) | **$^{81}$Kr** (($^{81}$Kr/Kr)$_{sample}$/ ($^{81}$Kr/Kr)$_{air}$) | **Cont[a]** (%) | **$^{39}$Ar$_{corr}$** (%modern) | **$^{81}$Kr$_{corr}$** (($^{81}$Kr/Kr)$_{sample}$/ ($^{81}$Kr/Kr)$_{air}$) |
|---|---|---|---|---|---|---|---|
| **1** | <0.62 | 17 ± 8 | 5.1 ± 0.5 | 0.19 ± 0.02 | <0.8 | 17 ± 8 | 0.19 ± 0.02 |
| **2** | 1.3 ± 0.4 | 8 ± 6 | 2.8 ± 0.1 | 0.30 ± 0.03 | 1.7 ± 0.5 | <12 | 0.29 ± 0.03 |
| **3** | <1.23 | <10 | <0.5 | 0.06 ± 0.02 | <1.6 | <10 | 0.06 ± 0.02 |
| **4** | 4.6 ± 0.5 | 19 ± 11 | 11.8 ± 0.4 | 0.06 ± 0.02 | 5.9 ± 0.6 | 12 ± 11 | 0.00 ± 0.02 |
| **5** | <0.86 | <8 | <0.5 | 0.38 ± 0.03 | <1.1 | <8 | 0.38 ± 0.03 |
| **6** | 9.4 ± 0.6 | n.m. | <0.5 | 0.12 ± 0.02 | 11.9 ± 0.7 | n.m. | 0.00 ± 0.02 |
| **7*** | <1.02 | <14 | <0.5 | 0.03 ± 0.02 | <1.3 | <14 | 0.03 ± 0.02 |

[a] calculated based on $^{85}$Kr. See Electronic Appendix A for details.

n.m. not measured

Table 5: Modeled mineral saturation states (SI) and partial pressure of $CO_2$ obtained from measured data and at calcite equilibrium. Model calculations were performed with PHREEQC using the Wateq4f database for wells 1 and 2 and the Pitzer database for the other wells (no data for quartz and barite).

| Well | pH field | $CO_2$ from field data (mmol/kg) | log $pCO_2$ from field data | pH adjusted | $CO_2$ adjusted (mmol/L) | log $pCO_2$ at calcite equilibrium | SI Calcite | SI Dolomite | SI Gypsum | SI Celestite | SI Barite | SI Quartz |
|---|---|---|---|---|---|---|---|---|---|---|---|---|
| **1** | 7.22 | 3.12 | -2.19 | 6.85 | 3.46 | -1.82 | 0.00 | -0.10 | -1.05 | -1.00 | 0.23 | 0.15 |
| **2** | 7.52 | 4.46 | -2.23 | 7.20 | 4.73 | -1.91 | 0.00 | -0.05 | -1.86 | -1.81 | 0.06 | 0.11 |
| **3** | 8.10[a] | 0.47 | -4.11 | 6.55 | 1.07 | -2.12 | 0.00 | 0.09 | -0.11 | -0.09 | – | – |
| **4** | 7.07 | 0.74 | -2.75 | 6.61 | 0.89 | -2.25 | 0.00 | 0.08 | -0.15 | -0.16 | – | – |
| **5** | 7.55 | 2.70 | -2.62 | 6.80 | 3.18 | -1.84 | 0.00 | 0.10 | -0.12 | -0.16 | – | – |
| **6** | n.m. | 0.50 | -2.82 | 5.75 | 2.15 | -1.24 | 0.00 | -0.20 | -1.10 | -1.11 | – | – |
| **7*** | 5.74[a] | 1.51 | -2.35 | 6.36 | 2.06 | -1.68 | 0.00 | 0.03 | 0.01 | 0.07 | – | – |

[a] unreliable field measurements

Table 6: Proportions $f_i^{H2O}$ of the meteoric water, glacial meltwater, and brine end members in the deep groundwater of the CAS calculated according to the modified mixing model from Raidla et al. (2009). Based on the water budget, a noble gas budget is calculated with the proportions of Kr, $f_i^{Kr}$, calculated according to Equation 7. Also shown are results for selected samples from Northern Estonia (Raidla et al., 2009): Muuga Port (NE1), Kuressaare (NE2), and Ruhnu Island (NE3). More details on the calculation of the mixing proportions are provided in the Electronic Appendix B.

| | Well | 1 | 2 | 3 | 4 | 5 | 6 | 7* | NE1 | NE2 | NE3 |
|---|---|---|---|---|---|---|---|---|---|---|---|
| Water proportion | Meteoric | 0.63 ± 0.05 | 0.71 ± 0.05 | 0.15 ± 0.10 | 0.18 ± 0.10 | 0.71 ± 0.03 | 0.00 ± 0.00 | 0.38 ± 0.08 | 0.20 ± 0.05 | 0.60 ± 0.05 | 0.68 ± 0.05 |
| | Glacial | 0.25 ± 0.05 | 0.25 ± 0.05 | 0.13 ± 0.06 | 0.10 ± 0.06 | 0.00 ± 0.02 | 0.00 ± 0.04 | 0.02 ± 0.04 | 0.80 ± 0.05 | 0.38 ± 0.05 | 0.20 ± 0.05 |
| | Brine | 0.12 ± 0.01 | 0.03 ± 0.00 | 0.72 ± 0.06 | 0.71 ± 0.05 | 0.28 ± 0.02 | 1.00 ± 0.04 | 0.60 ± 0.05 | 0.00 ± 0.00 | 0.02 ± 0.00 | 0.13 ± 0.01 |
| Kr proportion | Meteoric | 0.58 ± 0.08 | 0.62 ± 0.07 | 0.29 ± 0.21 | 0.38 ± 0.22 | 0.93 ± 0.05 | 0.00 ± 0.00 | 0.74 ± 0.15 | 0.13 ± 0.04 | 0.47 ± 0.07 | 0.65 ± 0.08 |
| | Glacial | 0.40 ± 0.08 | 0.38 ± 0.07 | 0.46 ± 0.19 | 0.37 ± 0.20 | 0.00 ± 0.05 | 0.02 ± 0.22 | 0.05 ± 0.13 | 0.87 ± 0.04 | 0.52 ± 0.07 | 0.33 ± 0.08 |
| | Brine | 0.02 ± 0.01 | 0.01 ± 0.00 | 0.25 ± 0.06 | 0.26 ± 0.06 | 0.07 ± 0.02 | 0.98 ± 0.22 | 0.21 ± 0.05 | 0.00 ± 0.00 | 0.00 ± 0.00 | 0.02 ± 0.01 |

Table 7: Groundwater ages deduced from $^{81}Kr$ using Equations 5 to 7.

| Well | $^{81}t$ | $^{81}t^{m}_{glacial-meteoric}$ (ka) | $t_{mix}$ |
|---|---|---|---|
| **1** | 548 ± 35 | 541 ± 28 | > 611 |
| **2** | 408 ± 35 | 407 ± 27 | > 411 |
| **3** | 929 ± 114 | 821 ± 89 | > 1'144 |
| **4** | >1'300 | >1'227 | > 1'279 |
| **5** | 319 ± 26 | 293 ± 21 | > 565 |
| **6** | >1'300 | >1'174 | > 1'300 |
| **7*** | 1'157 ± 266 | 1'067 ± 195 | > 1'152 |

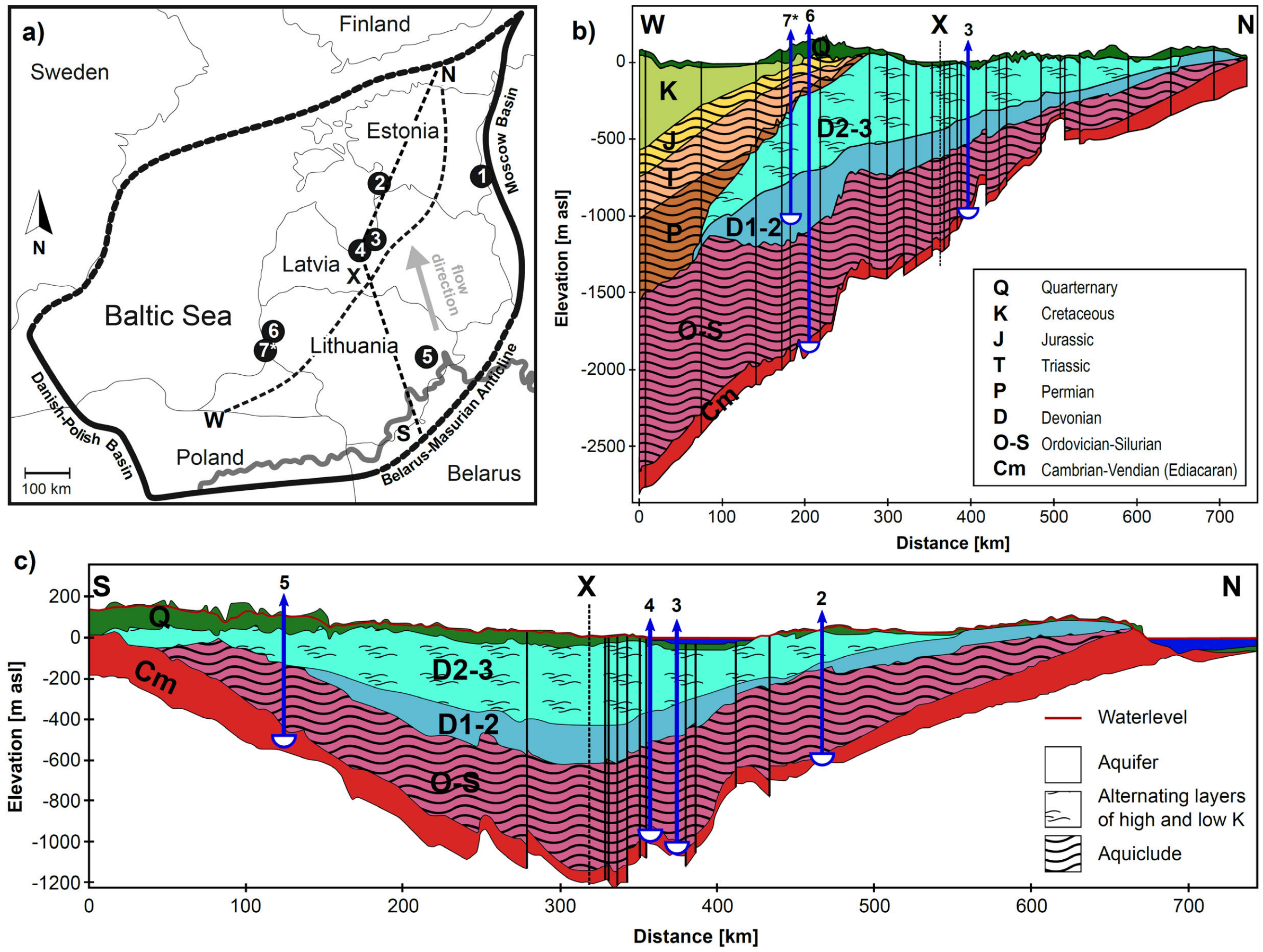

Figure 1: Geological setting of the Baltic Artesian Basin (BAB) and locations of the sampled wells (modified after Virbulis et al., 2013). Figure a) denotes the extent of the Cambrian aquifer system (CAS, thick black line) with the dashed parts indicating the aquifer outcrop areas. The thick gray line along the Belarus-Masurian Anticline shows the extent of the Scandinavian ice sheet during the last glacial maximum (Kalm, 2012). Figures b) and c) are cross-sections of the BAB along the fine dashed lines in 1a), showing the hydrogeological setting of the studied wells (200x vertical exaggeration). Vertical thin lines sketch the major faults. As a reminder that well 7* is the only well located in the overlying Devonian aquifer, it is marked with a star throughout the paper.

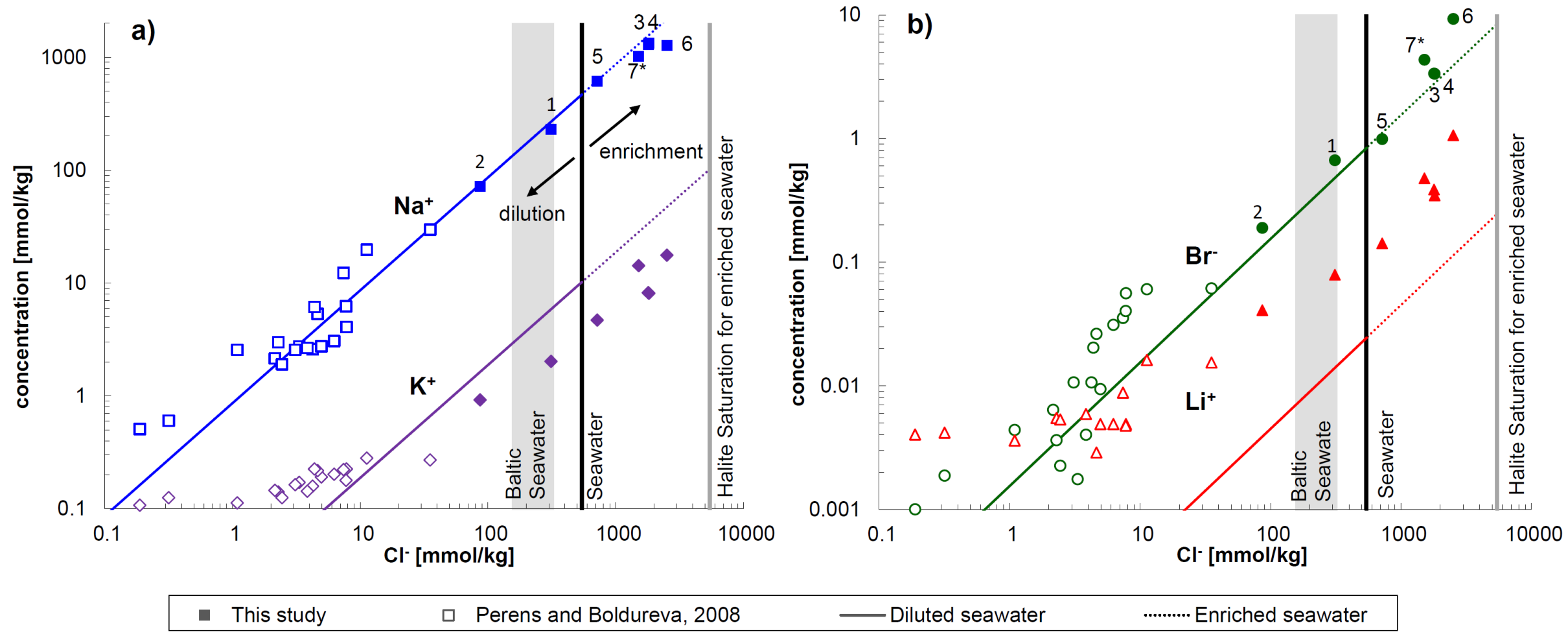

Figure 2: Concentrations of a) $Na^+$ and $K^+$ and b) $Li^+$ and $Br^-$ versus $Cl^-$. Solid symbols represent the groundwater samples of this study and open symbols are samples from Northern Estonia (data from Perens and Boldureva, 2008). The lines represent seawater that was diluted with pure water (solid lines) and enriched by evaporation (dotted lines) assuming present-day average seawater composition.

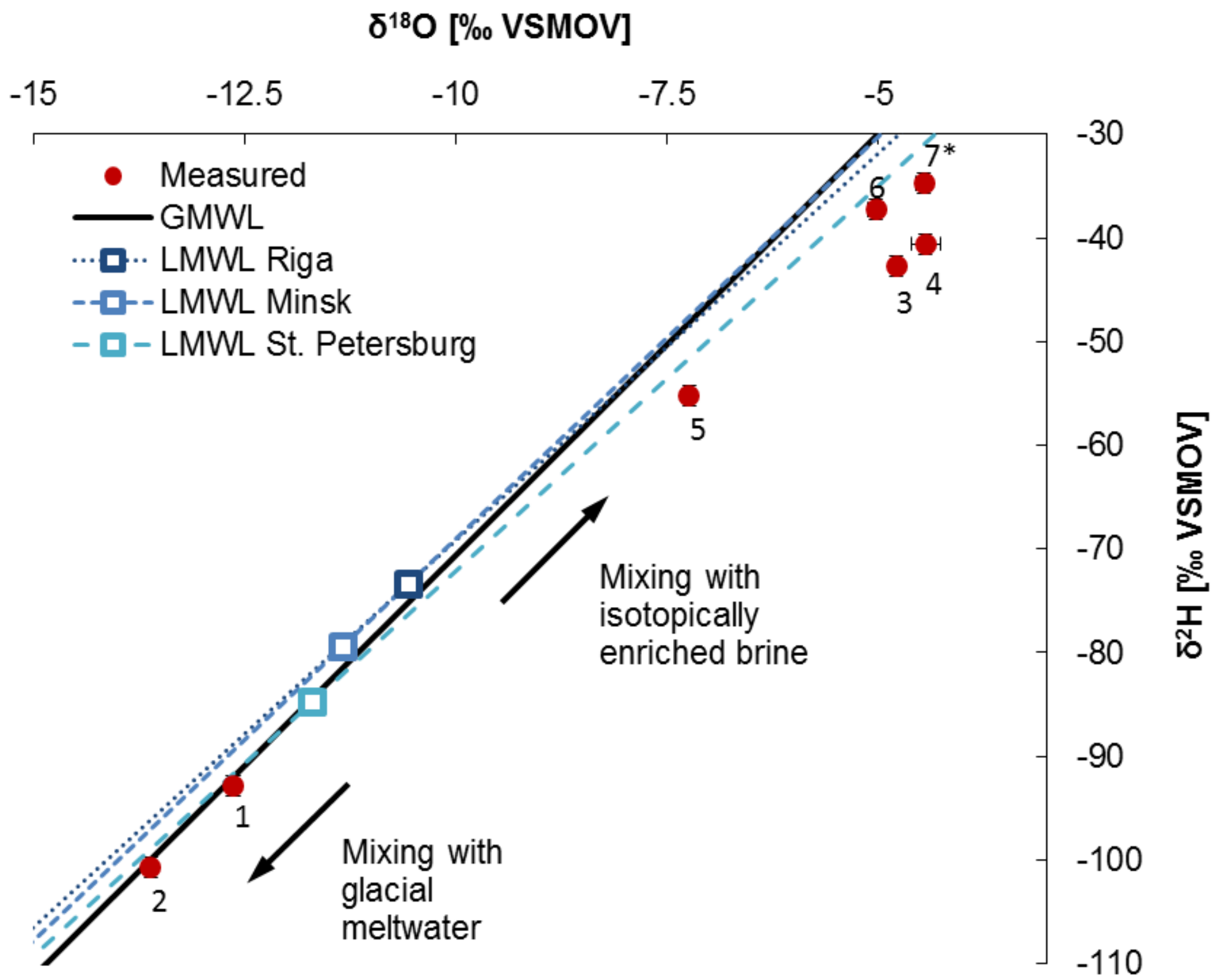

Figure 3: Relationship of $\delta^{18}O$ with $\delta^{2}H$, showing the discrimination between brackish and saline groundwater in the CAS. Local meteoric water lines (LMWL) and precipitation-weighted averages (squares) are based on data from the nearest GNIP stations (IAEA/WMO, 2016).

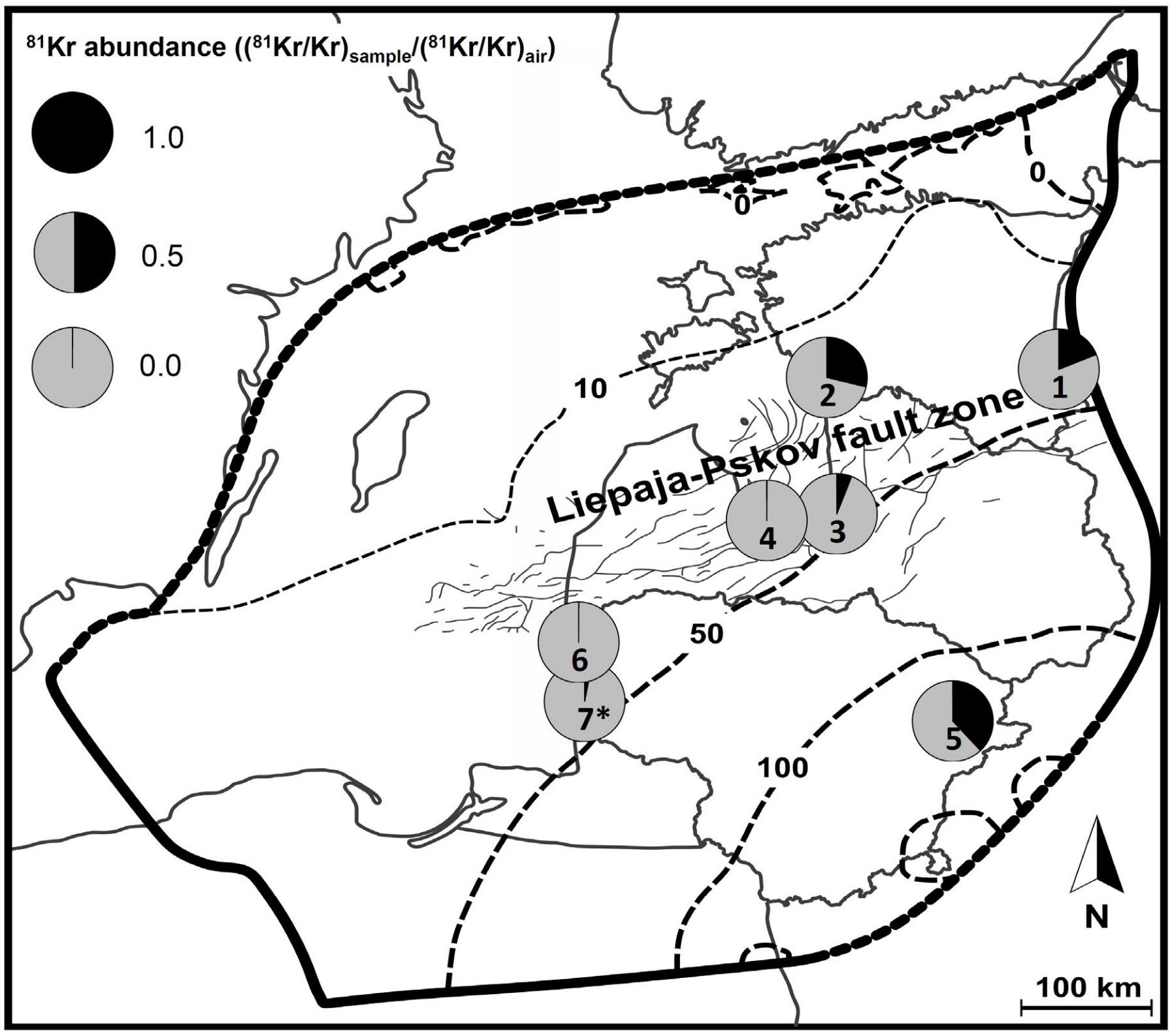

Figure 4: The spatial pattern of contamination-corrected $^{81}$Kr abundances. Dashed black lines are isobars of the present-day piezometric head (m asl) in the CAS modelled by Virbulis et al. (2013).

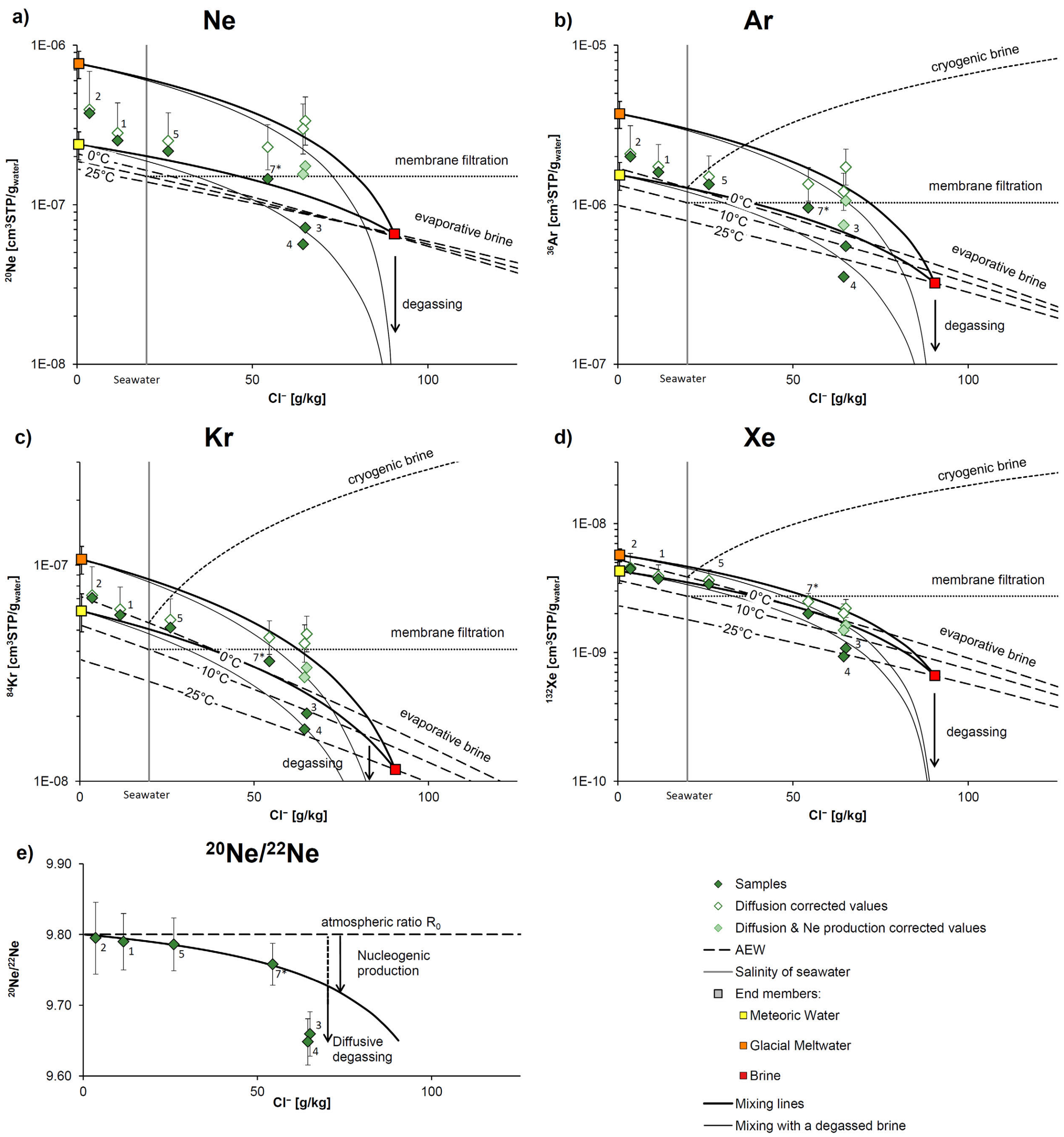

Figure 5: a)–d) Measured concentrations (solid diamonds) of atmosphere-derived noble gas isotopes and e) Ne isotopic ratios versus $Cl^-$, which is a proxy for the brine proportion. The evolution of noble gas concentrations during brine genesis depends strongly on the brine formation process (dashed and dotted black lines). Noble gas concentrations for air-equilibrated water (long dashed lines) are calculated according to Smith and Kennedy (1983) for various temperatures. Solid black lines in a) to d) show the range of concentrations expected for a mixing of the three end members (squares, numerical values of the noble gas signatures for the three end members are listed in the Electronic Appendix B) with and without degassing prior to mixing. Noble gas concentrations corrected for degassing according to Equations 3 and 4 (based on the $^{20}Ne/^{22}Ne$ ratio and assuming the degassing happened after mixing) are depicted as open diamonds. Taking into account the possibility of nucleogenic production of $^{20}Ne$ and $^{22}Ne$ for the brine end member results in the mixing line shown in e). In this case, the degassing correction for samples 3 and 4 in a) to d) is smaller and degassing is negligible for the rest of the samples.

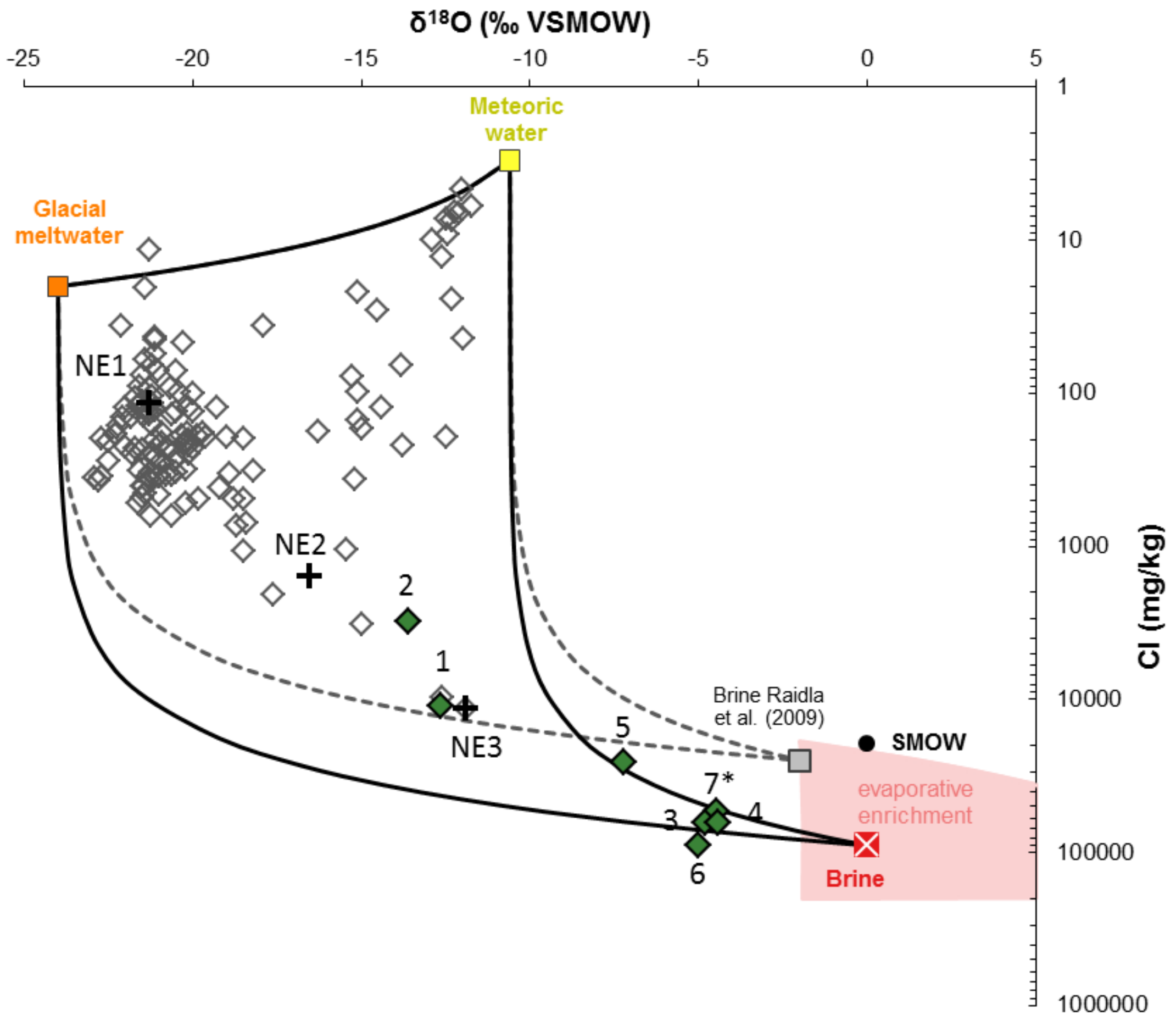

Figure 6: Relation between $Cl^-$ and $\delta^{18}O$, allowing the identification of end members (EM) based on the mixing model proposed by Raidla et al. (2009). The new data (filled symbols) result in an adjustment of the brine end member compared to that proposed by Raidla et al. (2009) based on samples from Northern Estonia (open symbols), without interfering with the proposed general mixing behavior of the less saline groundwaters. Also shown are the standard mean ocean water (SMOW) and the compositional range of evaporating paleo seawater. The change in $\delta^{18}O$ during evaporative enrichment is calculated using the equation provided in Ferronsky (2015) for an evaporating reservoir without recharge at 25°C, with the kinetic fractionation calculated according to Gat et al. (2001). The shaded area represents a range of atmospheric conditions (isotopic

composition of atmospheric water vapor $\delta_a$ and relative humidity h) during evaporation from $\delta_a$=−20‰ and h = 0.65 to $\delta_a$=−13‰ and h = 0.5 (corresponding to conditions reported for the Mediterranean (Gat et al., 2003). The measurement uncertainty is generally smaller than the symbol size. Numerical values of the $Cl^-$ and $\delta^{18}O$ signatures for the three end members are listed in the Electronic Appendix B.

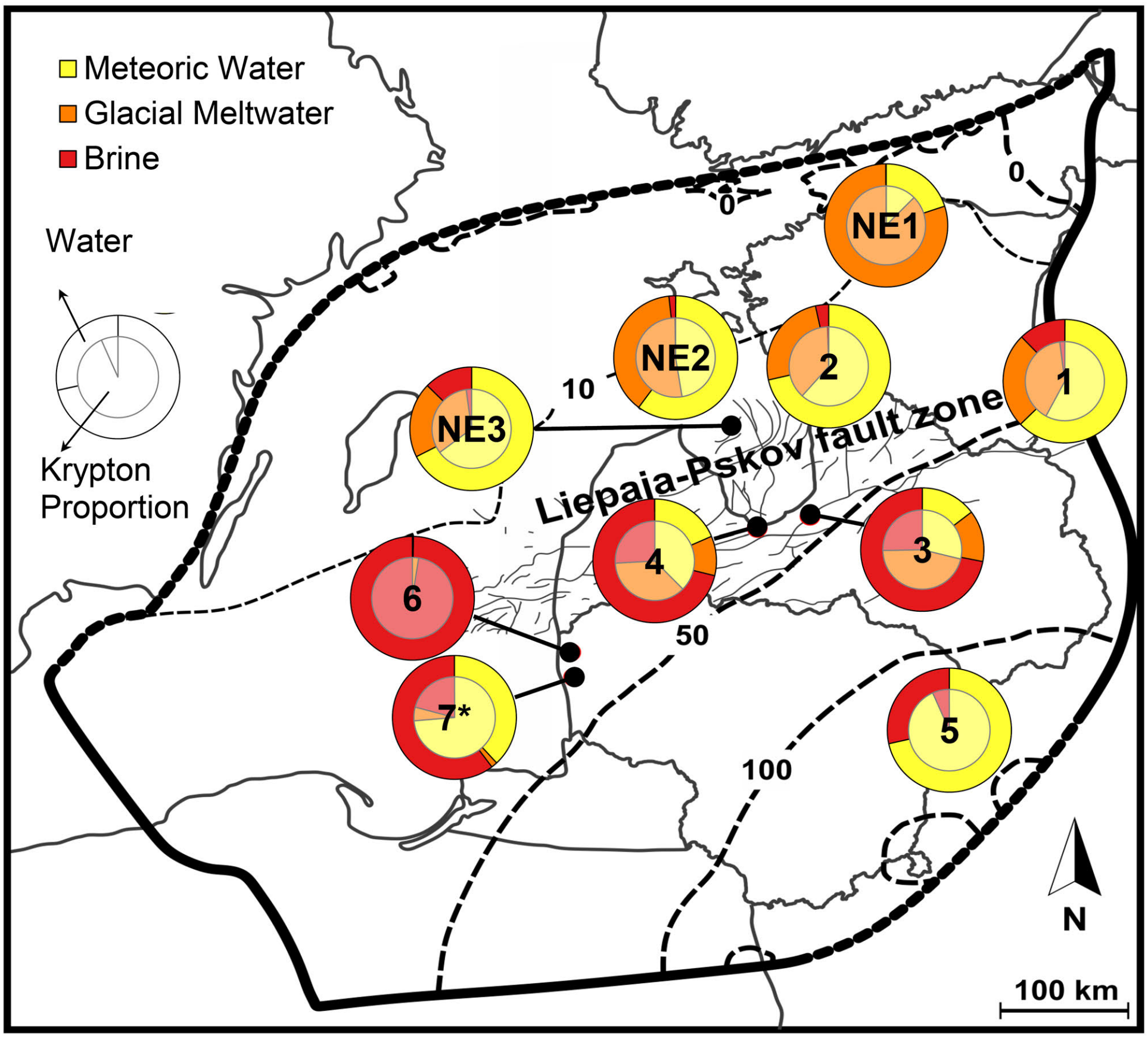

Figure 7: Spatial distribution of the water mixing proportions of the three end members in the deep groundwater of the CAS and the corresponding proportions of Kr. Present-day hydraulic heads (dashed lines, in m asl) in the CAS are according to Virbulis et al. (2013). Also shown are results for selected samples from Northern Estonia (NE1–NE3, data from Raidla et al., 2009).

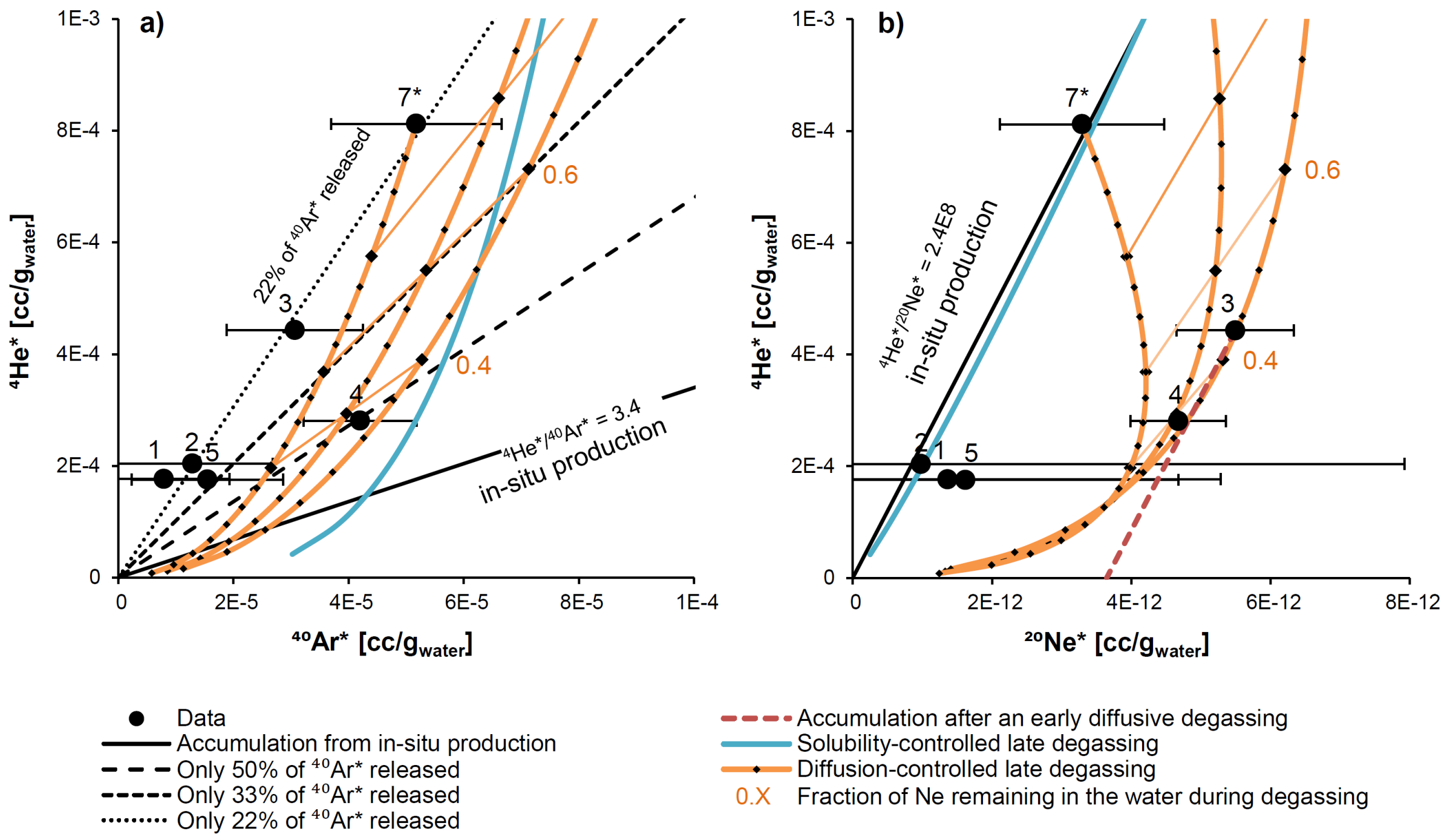

Figure 8: Radiogenic $^{4}He^{*}$ and $^{40}Ar^{*}$ and nucleogenic $^{20}Ne^{*}$ concentrations and calculated accumulation lines for in-situ production and partial release of $^{40}Ar^{*}$ into the water phase (black lines). Orange lines and the blue line represent the apparent changes in $^{4}He^{*}$, $^{20}Ne^{*}$, and $^{40}Ar^{*}$ resulting from a late degassing (diffusion-controlled and solubility-controlled, respectively), starting from sample 7 or from hypothetical samples with the same accumulation ratio but 50% and 100% higher concentrations. The thin orange lines and numbers refer to the remaining Ne fraction in the water during the course of the degassing (e.g. 40% to 45% of initial Ne in b) for wells 3 and 4). An early degassing and the subsequent accumulation of $^{4}He^{*}$, $^{20}Ne^{*}$, and $^{40}Ar^{*}$ is shown by the red dashed line in b). An early diffusive degassing shifts the $^{20}Ne/^{22}Ne$ ratio away from the atmospheric ratio, resulting in a significant amount of apparent $^{20}Ne^{*}$ before accumulation begins.

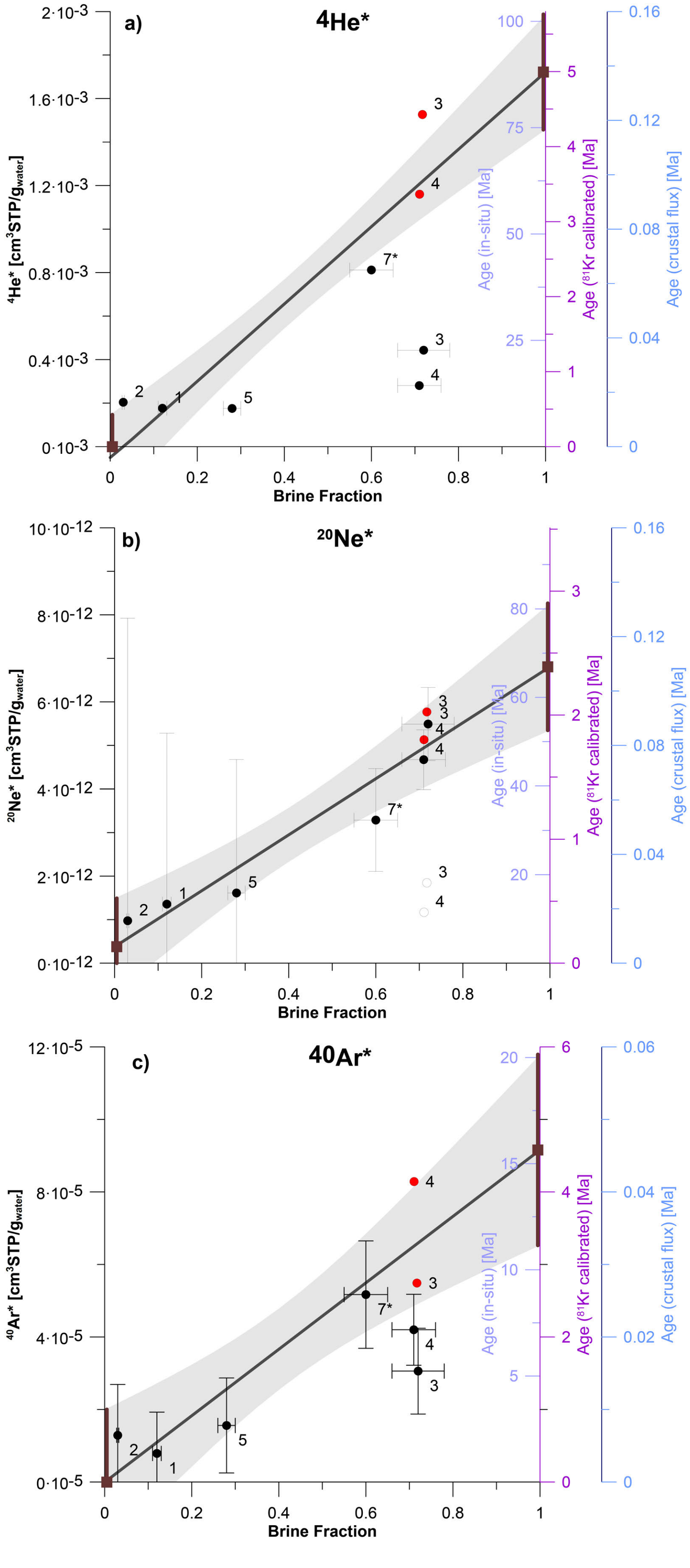

Figure 9: Concentrations of a) $^{4}He^{*}$, b) $^{20}Ne^{*}$, and c) $^{40}Ar^{*}$ as a function of the fraction of brine water. Black symbols refer to uncorrected values whereas open symbols are concentrations corrected for an early degassing (saline samples only) and red symbols are concentrations corrected for a late degassing as determined from Fig. 8. Based on a linear fit of the values corrected for early degassing (black line, the gray shading indicates the 1σ-uncertainty), concentrations in pure brine and the meteoric/glacial components are estimated (brown squares). On the right side, age scales for in-situ production, an accumulation rate based on $^{81}Kr$ ages of wells 1 and 2, and an external crustal flux are shown.

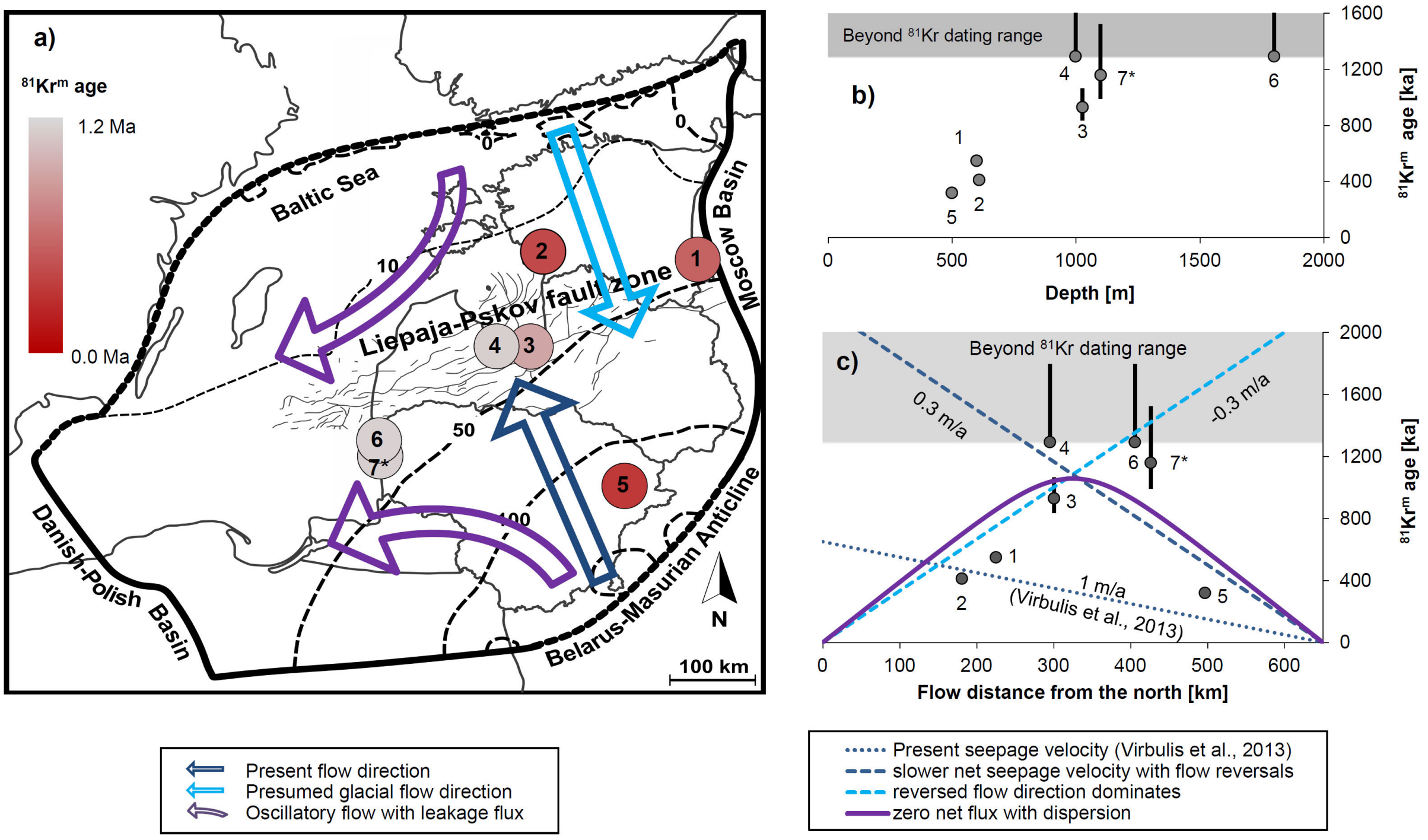

Figure 10: a) spatial distribution of $^{81}Kr^{m}$ ages of the meteoric water and glacial meltwater component. The arrows indicate different flow patterns (explained in the text). b) $^{81}Kr^{m}$ ages of the glacial and meteoric component versus depth. c) $^{81}Kr^{m}$ ages of the glacial and meteoric component as a function of flow distance from the Baltic Sea towards the Belarus-Masurian Anticline. The dashed lines represent groundwater ages corresponding to different net seepage velocities in both directions. The solid line represents ages corresponding to calculated $^{81}Kr$ concentrations for zero net flux at an average seepage velocity modulus of 2 m/a and a dispersivity of 10 km.

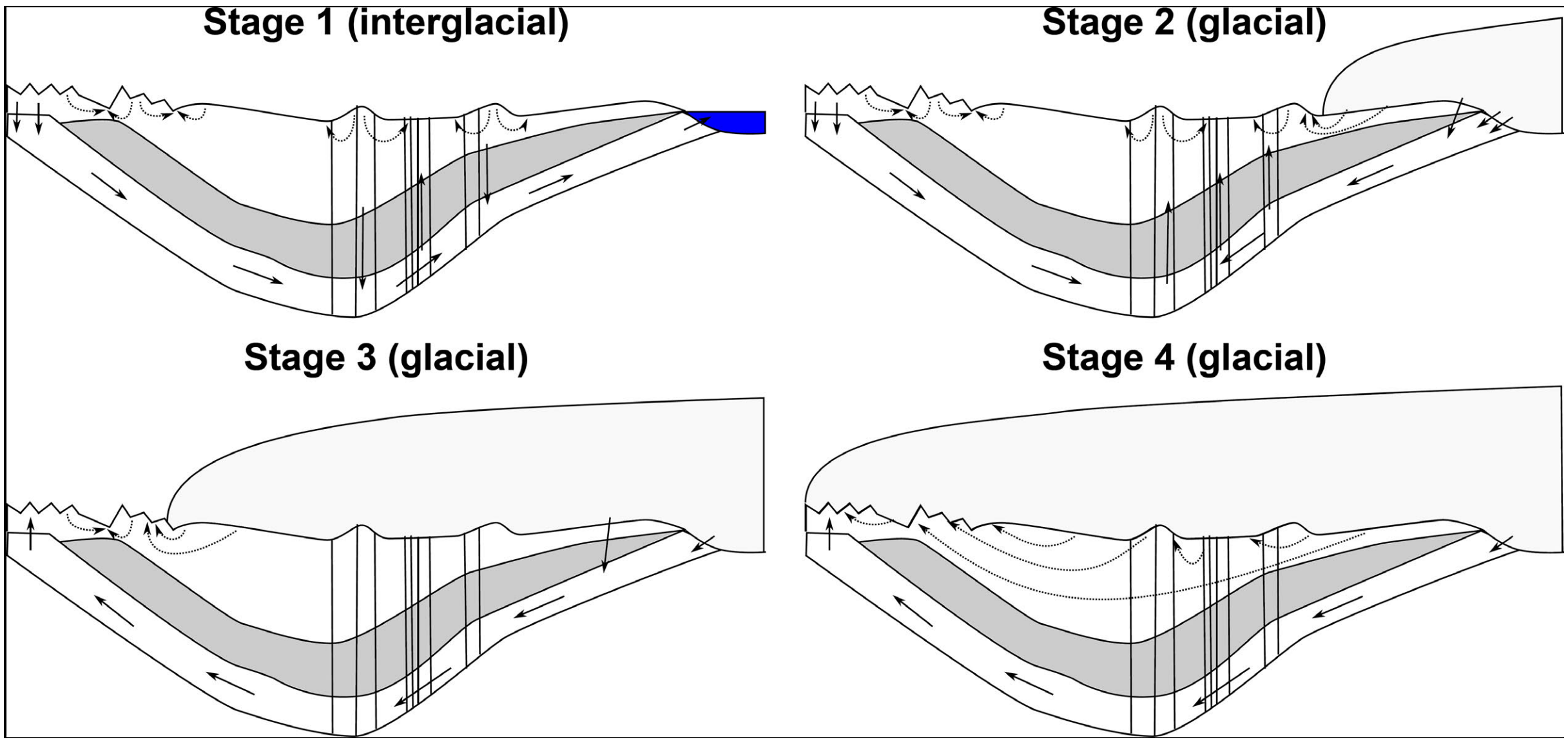

Figure 11: Presumed flow patterns in interglacial and glacial periods, controlled by the expansion and demise of the ice sheet. In the shallow aquifers (dashed arrows), and the fault zone, flow is mainly driven by topography during interglacial periods. Once covered by ice, we assume that the fault zone has negligible influence on water flows and flow is mainly driven by ice sheet topography.

# Supplementary Material

## A. Fieldwork and Contamination Correction

The seven sampling locations of this study (Fig. 1a, Table 1) were selected based on low radiocarbon activities from previous campaigns. Four of the wells (1–4) are artesian; otherwise, submersible pumps were used for sampling. During sampling, any water treatment (e.g. any admixture of additives) in actively used wells was paused.

Contamination from leakage could occur in the well due to a corroded well casing or in the gas extraction or noble gas separation system, although the extraction system was tested for leaks prior to and during sampling. Contamination detection is based on $^{85}$Kr, since the deep waters are presumably older than 60 years and thus $^{85}$Kr-free (Sturchio et al., 2014):

$$m = \, ^{85}Kr_{measured} / ^{85}Kr_{cont} \tag{A1}$$

where *m* is the fraction of contaminant (water or air) and $^{85}Kr_{cont}$ the $^{85}$Kr value of the contaminant (78.5±6.7 dpm/cc$_{Kr}$ in air from Tallinn taken during campaign II). Contamination-corrected $^{81}$Kr and $^{39}$Ar values are given by:

$$^{81}Kr_{corr} = \frac{1}{1-m}\left(^{81}Kr_{measured} - m\right) \quad \text{and} \quad ^{39}Ar_{corr} = \frac{1}{1-m}\left(^{39}Ar_{measured} - m\right) \tag{A2}$$

Assuming that $^{81}Kr_{cont}$ = 1 and $^{39}Ar_{cont}$ = 100%modern. The estimated contamination for each well is given in Table 4. Only well 6 seems to be substantially contaminated (12%). Contamination with atmospheric air during sampling only affects the gas tracers of campaign II ($^{85}$Kr, $^{81}$Kr, and $^{39}$Ar). Intrusion of relatively young water, on the other hand, would also affect all other parameters measured in the different campaigns. Contamination by shallow groundwater through a leaky well casing could be underestimated, depending on the groundwater residence time: If the water is very young, $^{85}Kr_{cont}$ is similar to the atmospheric 78.5 dpm/cc$_{Kr}$. For older water, some of the $^{85}$Kr will have decayed already, resulting in an underestimation of the admixed fraction of young water. This underestimation results in an overestimation of the true $^{81}$Kr abundance. Because it is difficult to separate the two processes and because there is no simple proxy of contamination for the other sampling campaigns, only $^{81}$Kr and $^{39}$Ar data were corrected.

## B. Mixing Calculations

The mixing proportions given in Table 6 in the main text were determined based on mass conservation for $\delta^{18}$O and $Cl^-$ and end member signatures as listed in Table A1. A Monte-Carlo simulation was performed (n=10’000) to determine the uncertainties of the calculated mixing proportions. The end member compositions for δ¹⁸O, Cl⁻, and ⁸¹Kr were assumed to be uniformly distributed within the ranges given in Table A1. For the measurements, normally distributed uncertainties with a standard deviation of 0.5‰ VSMOW, 5%, and values as reported in Table 4 were used for $\delta^{18}O$, $Cl^-$, and $^{81}Kr$, respectively. For end-member proportions close to 0 or 1, the median is more informative than the mean and therefore, we report the median of the Monte-Carlo simulation.

*Table A1: Cl-$\delta^{18}O$ and noble gas signatures of the end members. For $Cl^-$, $\delta^{18}O$, and $^{84}Kr$, which are used in the mixing calculations, the range of uncertainty is also given. These are based on the absolute uncertainty of the measurements for $Cl^-$ and $\delta^{18}O$ and on the standard deviation of $^{84}Kr$ concentrations of the glacial-meltwater-rich samples from Northern Estonia (Weissbach, 2014) for the glacial $^{84}Kr$.*

| | $Cl^-$ | $\delta^{18}O$ | $^{20}Ne$ | $^{36}Ar$ | $^{84}Kr$ | $^{132}Xe$ |
|---|---|---|---|---|---|---|
| | [mg/L] | [‰ SMOW] | (cm³STP/g) | | | |
| **Interglacial water** | 3 ± 10% | -10.6 ±1 | 2.4E-7 | 1.5E-6 | 6.1E-8 ±20% | 4.3E-9 |
| **Glacial meltwater** | 20 ± 10% | -24 ±1 | 7.7E-7 | 3.7E-6 | 1.1E-7 ±11% | 5.7E-9 |
| **Brine (undegassed)** | 90000 ± 10% | 0 ±1 | 6.6E-8 | 3.2E-7 | 1.1E-8 ±50% | 6.6E-10 |

The median and standard deviation of Kr proportions calculated from the mixing proportions and Kr concentrations of the end members (cf. Equation 7, Table A1) are also reported in Table 6. The calculated total Kr concentrations for each sample agree well with the measured Kr concentrations (Fig. A1) confirming the suitability of our mixing model for the noble gases. The $R^2$-value in Fig. A1 shows that 81% of the variation in noble gas concentrations can be explained by the three component mixing model (without samples 3 and 4 it is over 90%).

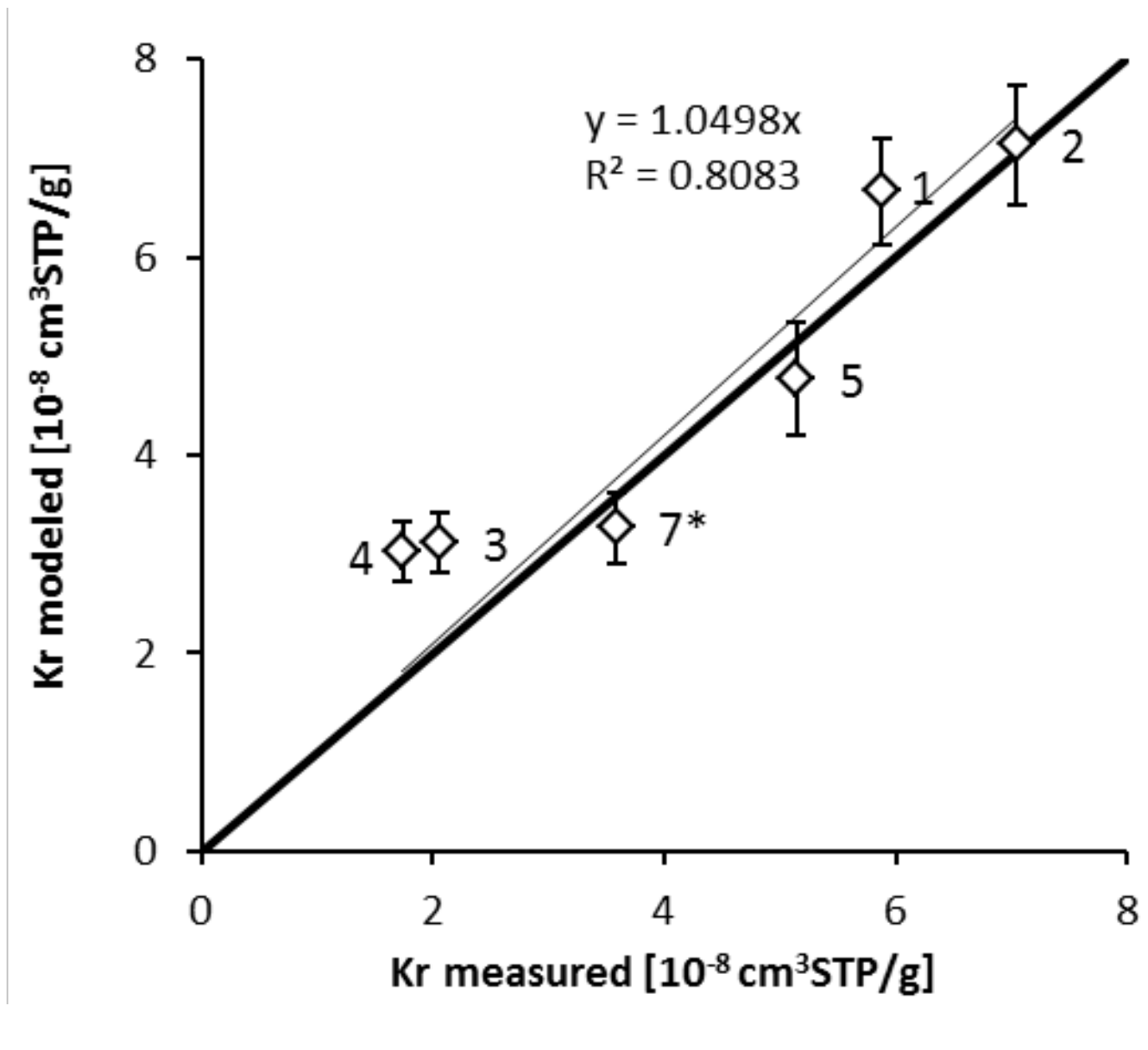

*Figure A1: Measured against modeled Kr and the 1:1 line.*

## C. Comparison of Ar concentrations

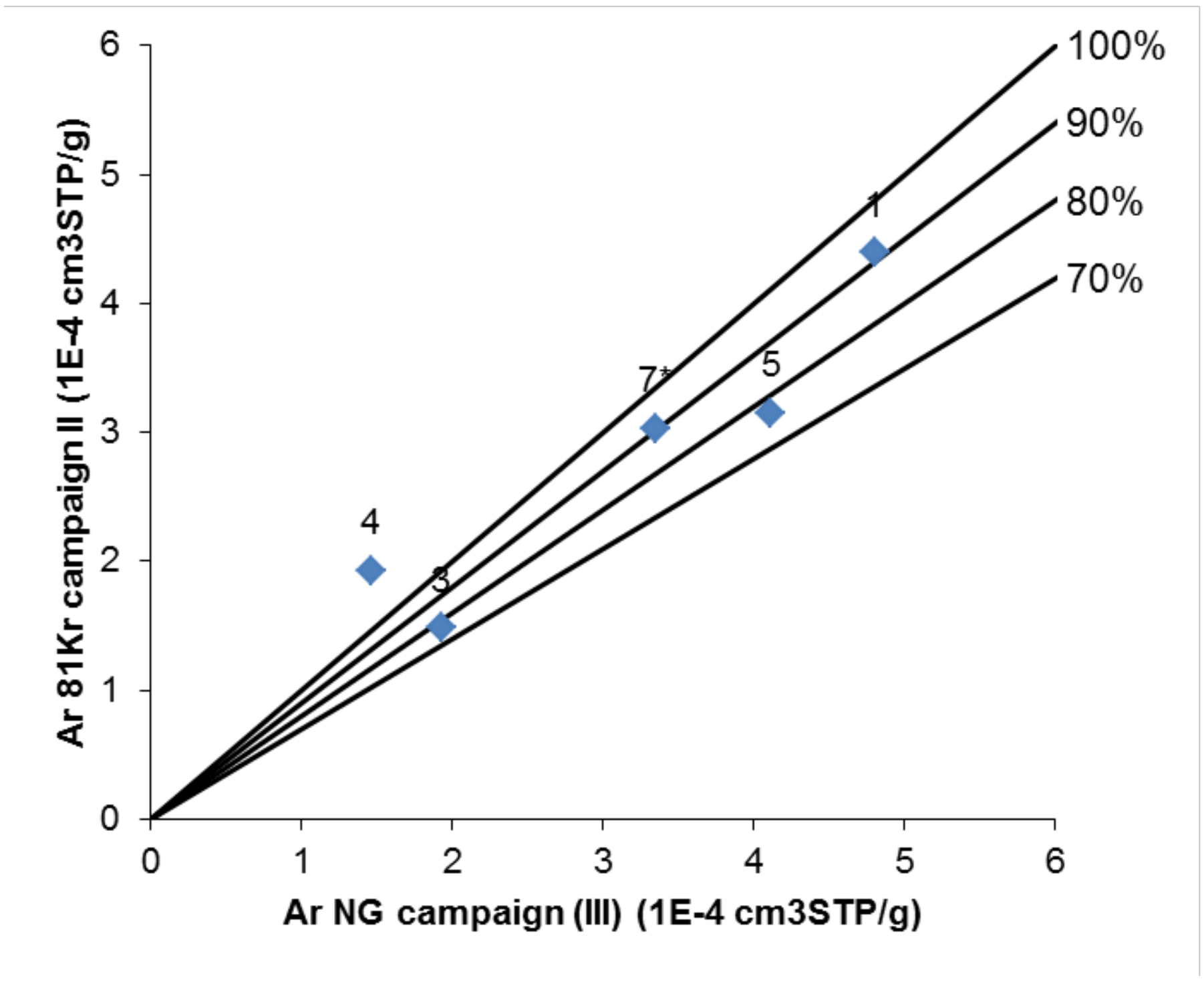

*Figure A2: Ar concentration measured from the Cu tubes collected during campaign III versus Ar concentrations inferred from the extracted gas for $^{81}$Kr dating during campaign II.*